\documentclass[10pt,journal,compsoc]{IEEEtran}

\usepackage[cmex10]{amsmath}

\usepackage{tikz}
\usepackage{adjustbox}
\usepackage{algorithm, algorithmic}             % format of the algorithm
\usepackage{booktabs}
\usepackage{epstopdf}
\usepackage{enumerate}
\usepackage{pdfpages}
\usepackage{afterpage}
\usepackage{color, colortbl}
\usepackage{footmisc}
\usepackage{etoolbox}
\usepackage{tablefootnote}
\usepackage{eucal}

\usepackage{xcolor}

\usepackage{mathtools}
\DeclarePairedDelimiter{\ceil}{\lceil}{\rceil}
\DeclarePairedDelimiter\floor{\lfloor}{\rfloor}
\usepackage{mathrsfs}  
\usepackage{pifont}% http://ctan.org/pkg/pifont
\newcommand{\cmark}{\ding{51}}%
\newcommand{\xmark}{\ding{55}}%
\usetikzlibrary{patterns}

\ifCLASSOPTIONcompsoc
\usepackage[nocompress]{cite}
\else
\usepackage{cite}
\fi

\usepackage{enumitem}
\setlist{nosep}

\usepackage{amsthm}
\theoremstyle{definition}

\usepackage{graphicx}
\graphicspath{{./figures/}}

\usepackage[pdftex]{hyperref}

\usepackage{array}
\makeatletter
\newcommand{\printfnsymbol}[1]{%
	\textsuperscript{\@fnsymbol{#1}}%
}
\makeatother

\ifCLASSOPTIONcompsoc
\usepackage[caption=false,font=footnotesize,labelfont=sf,textfont=sf]{subfig}
\else
\usepackage[caption=false,font=footnotesize]{subfig}
\fi

\usepackage{multirow}
\usepackage{url}

\usepackage{hyperref}

\usepackage{url}

\begin{document}

\title{Energy-aware Task Scheduling with Deadline Constraint in DVFS-enabled Heterogeneous Clusters}
	
\author{
	Xinxin~Mei, Qiang~Wang, Xiaowen~Chu\printfnsymbol{1}, Hai-Liu, Yiu-Wing~Leung, Zongpeng Li

	\IEEEcompsocitemizethanks{\IEEEcompsocthanksitem Xinxin Mei, Qiang Wang, Xiaowen Chu and Yiu-Wing Leung are with the Department
		of Computer Science, Hong Kong Baptist University. \protect\\E-mail: \{xxmei, qiangwang, chxw, ywleung\}@comp.hkbu.edu.hk} 
		
	\IEEEcompsocitemizethanks{\IEEEcompsocthanksitem Hai Liu is with the Department of Computing, Hang Seng University of Hong Kong. \protect\\E-mail:{hliu}@hsu.edu.hk}
		
	\IEEEcompsocitemizethanks{\IEEEcompsocthanksitem Zongpeng Li is with the Department of Computer Science, the University of Calgary. \protect\\E-mail:{zongpeng}@ucalgary.ca}
		
	\IEEEcompsocitemizethanks{\IEEEcompsocthanksitem Xinxin Mei and Qiang Wang contributed equally to this work. Xiaowen Chu is the corresponding author.}
		
	\IEEEcompsocitemizethanks{\IEEEcompsocthanksitem This manuscript is an extension of the INFOCOM 2017 paper entitled "Energy efficient real-time task scheduling on CPU-GPU hybrid clusters".}	
}

\IEEEtitleabstractindextext{%
	\begin{abstract}
		Energy conservation of large data centers for high performance computing workloads, such as deep learning with big data, is of critical significance, where cutting down a few percent of electricity translates into million-dollar savings. This work studies energy conservation on emerging CPU-GPU hybrid clusters through dynamic voltage and frequency scaling (DVFS). We aim at minimizing the total energy consumption of processing a batch of offline tasks or a sequence of real-time tasks under deadline constraints. We derive a fast and accurate analytical model to compute the appropriate voltage/frequency setting for each task, and assign multiple tasks to the cluster with heuristic scheduling algorithms. In particular, our model stresses the nonlinear relationship between task execution time and processor speed for GPU-accelerated applications, for more accurately capturing real-world GPU energy consumption. In performance evaluation driven by real-world power measurement traces, our scheduling algorithm shows comparable energy savings to the theoretical upper bound. With a GPU scaling interval where analytically at most 36\% of energy can be saved, we record 33-35\% of energy savings. Our results are applicable to energy management on modern heterogeneous clusters.
	\end{abstract}
		
}
	
\maketitle
\IEEEdisplaynontitleabstractindextext

\IEEEpeerreviewmaketitle
	
\section{Introduction}\label{sec:introduction}
Energy conservation and power management is now a major subject of study in high performance computing platforms. It aims in reducing the energy consumed by computer systems while maintaining a good level of quality of service (QoS). 
As the demand for computation and data processing is growing exponentially, especially with the rapid development of deep learning techniques, the high-performance distributed systems equipped with many-core accelerators (such as GPUs, Intel MICs, and FPGAs) are becoming more and more indispensable in many academic and industrial applications. However, the cost to power those systems during their lifetime surpasses that to manufacture them. For example, the GPU-accelerated DeepMind computing center, best known for defeating professional human players in the strategy board game of Go, was acquired for about 600 million dollars, while its annual electricity bill is pegged at 150 million \cite{googleAI2016,silver2016mastering}. Besides, the advanced language model GPT-3 \cite{floridi2020gpt}, designed with over 150 billion parameters to generate human-like texts, needs at least 4.6 million dollars of training cost.
Moreover, training deep learning models and recent automated machine learning (autoML) \cite{xin2021automl} techniques require a lot of trials and errors that would probably increase the cost several-fold.
Even saving a few percentages of the energy consumed in such computing centers can bring tremendous financial gains.

Graphical processing units (GPUs) have become prevalent and necessary accelerators in modern data centers, especially those for deep model training. Compared to the mainstream CPUs, modern GPUs are always around one order of magnitude faster and more power-efficient in terms of Flops-per-Watt \cite{powerS2016,survey2017}. In the TOP-500 supercomputer list \cite{top500} as of November 2020, 149 are equipped with accelerators, and 140 out of them are equipped with GPUs. Even in the Green-500 List that ranks supercomputers by performance per Watt, 8 out of the top 10 most energy-efficient supercomputers are accelerated with GPUs \cite{green500}.

Though the hybrid CPU-GPU clusters can achieve higher energy efficiency, their energy consumption is still very high. E.g., a single DGX A100 server from Nvidia consumes up to 6,500 Watts, nearly 50\% of which come from its 8 GPUs. To power a large-scale cluster remains a major item of expense for data centers, and the energy efficiency of GPU-accelerated clusters is an important direction of research due to the complicated relationship between the runtime performance and power consumption \cite{mei2014benchmark,mei2017tpds,tang2019dvfs_dl}. Despite the growing need for energy conservation on GPU-accelerated hybrid clusters, GPU energy management techniques only start to witness developments.

Two commonly used techniques for saving energy in data centers are dynamic voltage and frequency scaling (DVFS) and dynamic resource sleep (DRS). DVFS refers to the capability of adjusting the voltage and frequency of a processor dynamically, while DRS puts idle servers into deep sleep states (or simply turning them off) to conserve energy \cite{fan2007,isci2006}. However, simply transplanting CPU DVFS strategies onto GPU platforms could be ineffective \cite{ge2013,yuki2012}, mainly due to the following two reasons. First, most existing works on CPU DVFS only consider scaling the CPU voltage or frequency alone, while existing works have shown that the GPU core voltage, GPU core frequency, and GPU memory frequency are the major factors that affect the dynamic GPU power \cite{yuki2014,survey2017,wang2020dvfs}. Second, the execution time on a CPU is typically inversely proportional to the processor frequency, which is not always true on a GPU \cite{racing2015}. Many GPU-accelerated applications are memory-bounded and their performance is not only related to the GPU core frequency but also GPU memory frequency. Hence the tradeoff between the application execution time and its average power on GPUS becomes more complicated.

This article is an extension of our previous conference paper \cite{xin2017schedule}, which has explored how to efficiently assign multiple online tasks to the cluster with heuristic scheduling algorithms and GPU DVFS techniques. Based on that, we conduct a comprehensive study on energy efficient task mapping on CPU-GPU heterogeneous clusters with deadline constraints. Our major objective is to minimize the total energy consumption of executing either a batch of tasks (offline mode) or a sequence of real-time tasks (online mode), while guaranteeing the task deadlines. This requires both appropriate GPU voltage/frequency configuration and task scheduling strategy. Such a problem is of practical significance in the resource management of data centers \cite{anton2012,tang2016energy}. To tackle the energy minimization problem, we need to accurately understand the GPU performance model and power model. Towards this end, we consider three scaling variables that have significant impact on task execution time and power consumption: GPU core voltage, GPU core frequency, and GPU memory frequency. We first introduce our GPU performance and power model, which captures the nonlinear relationship between task execution time and the GPU core/memory frequency. We then apply optimization techniques to compute the optimal voltage and frequency setting in terms of minimizing the energy consumption for a single task with and without deadline constraints. As for task scheduling and mapping on hybrid servers, we first propose the offline solution EDL to it. We then extend the offline EDL algorithm to the online problem, where we combine GPU DVFS and dynamic resource sleep (DRS). One challenge in this scheduling problem is to achieve a good balance between dynamic energy consumption (which prefers low voltage/frequency and long execution time) and static energy consumption (which prefers high voltage/frequency and low execution time). Since the optimal DVFS settings obtained in our first step do not consider static energy, we introduce another variable named readjustment factor to allow a non-optimal voltage/frequency setting for better task packing and hence less static energy consumption. Our major contributions in this work can be summarized as follows.

\begin{itemize}
    \item This work presents a fast and accurate analytical GPU-specific DVFS model, and the optimization solution for tuning the DVFS setting of a single task.
    \item We propose energy efficient task scheduling algorithms on hybrid CPU-GPU clusters for both the offline and online modes with the following features: (i) it exploits GPU DVFS to conserve energy consumption without violating task deadlines; (ii) it effectively packs a set of tasks on a number of servers to reduce static energy consumption; (iii) it intelligently adjusts the DVFS setting for better energy savings.
    \item We conduct real GPU experiments on a set of benchmark applications to understand how much energy can be saved by GPU DVFS. We then design extensive simulations based on our experimental data to evaluate the effectiveness of our scheduling algorithm. Our simulation results show that nearly 34\% of the energy consumption can be saved. We make the model implementation open-source\footnote{\url{https://github.com/HKBU-HPML/GPU-DVFS-Job-Schedule}} for reproducing our experimental results.
\end{itemize}

The rest of the paper is organized as follows. Section \ref{sec:related_work} reviews related work. Section \ref{sec:formulation} describes our GPU power and performance models considering DVFS, and formulates the energy optimization problem. Section \ref{sec:solution} presents our optimization techniques and scheduling algorithms. Section \ref{sec:experiment} presents simulation results. Finally, we conclude the paper in Section \ref{sec:conclusion}.
\section{Related Work}\label{sec:related_work}
\subsection{GPU DVFS Performance and Power Modeling}

To maximize the energy efficiency brought by GPU DVFS, some existing models \cite{fan2019predictable,dvfs2019,huang2019} generally adopted machine learning methods to clarify the application patterns and predict the effects of DVFS on performance and energy. However, as the demand of task scheduling can be diverse, it is necessary to precisely predict the performance and power of a GPU application under different DVFS settings. 

As for performance modeling, recent studies leveraged the profiling technique to collect the kernel instruction information and then utilized either data-driven methods or pipeline analysis to model the kernel execution time. For example, Ali et al. \cite{ali2015} applied principle component analysis (PCA) to select 12 most crucial profiled performance counters and then adopted an MLP model to fit the kernel execution time. Wu et al. \cite{wu2015gpgpu} developed a performance model based on pattern clustering and classification considering the effect of DVFS.
Besides, some research projects \cite{wong2010demystifying,mei2014benchmark,mei2017tpds} developed a benchmark suite to demystify those GPU micro-architecture parameters. 
Based on the basic hardware information, a series of studies \cite{hong2009analytical,song2013simplified,wu2015gpgpu,wang2020dvfs} proposed several analytical models to estimate different degrees of memory traffic and computational workload of a kernel, and then calculate the execution cycles according to different cases. Nath et al. \cite{nath2015crisp} and Qiang et al. \cite{wang2020dvfs} also theoretically discussed the effect of GPU DVFS on the kernel execution time, which is much different from CPU DVFS.

As for power modeling, some state-of-the-art models have achieved remarkable accuracy using machine learning techniques. Vignesh et al. \cite{Vignesh2016} proposed an instantaneous GPU power prediction model using statistical regression techniques with those pivotal performance counters. Bishwajit et al. \cite{Dutta2018} explored the performance of several machine-learning prediction techniques on the GPU power prediction problem at different DVFS settings. Recently, Jo$\tilde{a}$o et al. \cite{gpupower2018} designed a set of 83 carefully crafted micro-benchmarks to stress the main GPU components under different frequency settings, which decomposed the contribution of different parts to the final power. These work finally demonstrated decent accuracy and generalization across different GPU platforms and applications. 

Although the existing profiling based methods for performance and power estimation have achieved considerable accuracy, collecting those instruction information may introduce dozens to hundreds of time consumption compared to purely executing the kernel \cite{wang2020dvfs}, which is critical to make an instantaneous decision for online real-time task scheduling. To decrease the modeling overhead, we present a low-cost but accurate analytical GPU-specific DVFS model, which shrinks the parameter set and focuses on the sensitivity of performance and power to GPU DVFS.

\subsection{DVFS Scheduling Algorithm}
A rich body of theoretical studies model processor power consumption with a single variable, the processor speed, which can be controlled by varying the processor’s voltage and frequency. Yao et al. studied task scheduling on a single processor. They proved that for the offline problem, the optimal speed during task processing is a constant \cite{yao1995}. Aydin et al. and Albers et al. proved that the offline multiprocessor scheduling problem to minimize energy consumption while meeting the task deadline is NP-hard \cite{aydin2003,Albers2014}. Aydin et al. also proved that evenly distributing the workloads among the multiple processors can minimize the energy consumption \cite{aydin2003}. Hong et al. proved that for a multiprocessor system, there is no online optimal scheduler \cite{hong1992}. Irani et al. studied speed scaling along with DRS \cite{Irani2007}. They found that the algorithms with DRS perform similarly to those without DRS.

Gharaibeh et al. verified that a CPU-GPU hybrid cluster
can achieve better performance and energy efficiency than a typical CPU cluster, for extremely large real-world graphic applications \cite{Gharaibeh2013}. Liu et al. integrated CPU DVFS and GPU task migration on a heterogeneous cluster \cite{liu2011}. 
In their model, a task can be divided into a CPU-subtask and a GPU-subtask, and the execution of the two subtasks is asynchronous. The CPU voltage is scaled for better CPU-GPU load-balance. Liu et al. studied power-efficient online scheduling algorithms on CPU-GPU heterogeneous clusters \cite{Liu2012}. In their model, the task is allowed to execute on either one CPU processor or one GPU processor. 
They examined earliest-deadline-first (EDF) and first-fit (FF) heuristic scheduling algorithms. They conducted experiments on a CPU-GPU platform, but because of the difficulty to measure GPU power consumption of different voltage/frequency states, they calculated the data instead.
Xie et al. \cite{xie2017schedule} explored how to minimize the total energy consumption by properly scheduling the tasks to the processors without violating their deadlines. They proposed a variant of the HEFT algorithm and discussed the cases with and without using DVFS.
%Ali et al. \cite{ali2020cloud} discussed two different priority based methods, earliest deadline first (EDF) and RM, and exploited the application of RM algorithm for real-time task scheduling in cloud environment, which has not been employed previously. 
Deng et al. \cite{deng2020tc} proposed an improved cuckoo search algorithm combined with a heuristic modification strategy. Their DVFS model considered different types of CPU processors. 
%by means of improving resource utilization, prioritizing efficiency, load balance, 
Other recent studies \cite{vincent2017energy,connor2018icpp,salami2021toc} also employed the DVFS techniques to provide efficient task scheduling in terms of fairness and energy conservation.  

We compare some recent proposed methods with ours in Table \ref{tab:compare_exist}. Our work differs from existing ones in two aspects. First, the above literature, except our former conference version \cite{xin2017schedule}, all assumed that the processor execution speed is linearly proportional to the processor voltage/frequency (despite the findings in \cite{Nath2015,wang2020dvfs}), and the energy consumption is monotonically increasing in the scaling interval. Following these assumptions, the appropriate voltage or frequency level is determined by the processor workload \cite{aydin2001}. In contrast, our energy function can be non-monotonic in the voltage/frequency scaling interval, and the optimal voltage/frequency is more related to task properties than processor workload. Second, our work discusses the impact of GPU DVFS on both online and offline task scheduling, which is partially absent in the previous work.
%Furthermore, most existing work ignores the impact of scaling the memory frequency on power consumption and job execution time. We explicitly consider GPU memory frequency as one major factor in reducing the overall energy consumption.

\begin{table}
	\centering
	\caption{Comparison against the exisiting studies} \label{tab:compare_exist}
	\begin{tabular}{ccccc} \hline
		Paper & Online & Offline & GPU & Energy Model \\ 
		& Scheduling & Scheduling & DVFS & for DVFS \\ \hline
		G. Xie \cite{xie2017schedule} & \xmark & \cmark & \xmark & monotonic \\ \hline
		X. Mei \cite{xin2017schedule} & \cmark & \xmark & \cmark & non-monotonic \\ \hline
		B. Salami \cite{salami2021toc}	& \cmark & \xmark & \xmark & monotonic \\ \hline
		Our work & \cmark & \cmark & \cmark & non-monotonic \\ \hline
	\end{tabular}
\end{table}

\section{System Modeling and Problem Formulation} \label{sec:formulation}
\subsection{System Modeling}
\subsubsection{GPU DVFS Power and Performance Modeling}
In the typical heterogeneous data center, the GPUs are the co-processors of the CPUs. They have independent core processors and memory space. The voltage/frequency of both the GPU computing core and GPU memory system can be scaled. We denote the GPU core voltage, core frequency, memory voltage and memory frequency by $V^{Gc}, f^{Gc}, V^{Gm}, f^{Gm}$, respectively. We can model the GPU runtime power and execution time with the functions of these four variables. 

In our previous studies \cite{survey2017,wang2020dvfs}, we scaled all of the four adjustable parameters of GPUs across five architecture generations (from Fermi to Volta), and exhaustedly studied the impact of GPU DVFS on the system energy conservation (typically a platform equipped with one CPU and one GPU) with a set of over twenty applications taken from CUDA SDK \cite{cudasdk100} and Rodinia benchmark suite \cite{che2009rodinia}. We introduce some findings as follows. 
\begin{enumerate}
	\item Scaling GPU memory voltage nearly has no influence on the system energy, because the GPU memory voltage has a narrow adjustable range and the GPU memory power consumption only accounts for a small part of the whole system. Thus in the modeling, to simplify the GPU power and performance modeling, we do not consider the GPU memory voltage scaling, either. On the other hand, scaling the memory frequency influences the task execution time significantly and the runtime power considerably.
	\item The GPU maximum core frequency is proportional to the core voltage ($V^{Gc}$), but the frequency referred model in literature, i.e. $f^{Gc}=\beta V^{Gc}$, does not hold. We denote the $f^{Gc}_{\text{max}}-V^{Gc}$ relationship as $f^{Gc}_{\text{max}}=g_1(V^{Gc})$. We derive a specialized $g_1(V^{Gc})$ function for our experimental Pascal platform based on the measurement results, where $g_1(V^{Gc})$ is sublinear.
	\item The system energy consumption is influenced by both the GPU core and memory frequency. They affect the system energy mostly in the application execution time. According to our experiments, different applications show different degrees of sensitivity to even one kind of frequency scaling. 
\end{enumerate}
In reality, the voltage/frequency scalable range of different GPU products may vary. Without loss of generosity, we compute the normalized value of $f^{Gc}, V^{Gc}$ and $f^{Gm}$ instead of the real value.

We model the GPU runtime power in Equation \eqref{eq:power_model}, where $P^{G0}$ is the summation of the power consumption unrelated to the GPU voltage/frequency scaling; $V^{Gc}, f^{Gc}, f^{Gm}$ denote the GPU core voltage, GPU core frequency, and GPU memory frequency respectively. $\gamma$ and $c^G$ are constant coefficients that depend on the hardware and the application characteristics, indicating the sensitivity to memory frequency scaling and the core voltage/frequency scaling respectively \cite{hong2010}. In this work, the parameters in the power modeling of an application are derived from its measured average runtime power consumption.

\begin{align}
P^G = \mathscr{P}(V^{Gc}, f^{Gc}, f^{Gm})=P^{G0}+\gamma f^{Gm} + c^G(V^{Gc})^2 f^{Gc} \label{eq:power_model} 
\end{align}
Performance modeling of GPU DVFS is rather complex \cite{Nath2015,wang2018,wang2020dvfs}. In this work, we seek a first-order mathematical model with a concise form to simplify the subsequent analysis of task scheduling. We formulate the performance function ($\mathscr{T}$) of a GPU-accelerated application as shown in Eq. \eqref{eq:time_model} \cite{yuki2014}, where $D$ represents the component that is sensitive to GPU frequency scaling, and $t^0$ represents the other component in task execution time. $\delta$ is a constant factor that indicates the sensitivity of this application to GPU core frequency scaling. We can always adjust the value of $D$ and $\delta$ to model the sensitivity to GPU memory frequency scaling as $1-\delta$. With $\delta, t^0$ and $D$ set to different values, the model is capable of simulating the various DVFS effects of a variety of applications.
\begin{align}
    t = \mathscr{T}(V^{Gc}, f^{Gc}, f^{Gm})=D(\frac{\delta}{f^{Gc}}+\frac{1-\delta}{f^{Gm}})+t^0 \label{eq:time_model}
\end{align}

It is notable that $f^{Gc}$ and $V^{Gc}$ are correlated. For a fixed $V^{Gc}$, the maximum allowed core frequency ($f^{Gc}_{max}$) is
determined by $V^{Gc}$. We use $V^{Gc}=g_1(V^{Gc})$ to denote this
relationship, which has been shown to be sublinear in [29]. 

With the above GPU DVFS power and performance model,
the GPU energy ($E^G$) consumed to process one task is the product of the runtime power and the execution time, as shown in Equation \eqref{eq:e_model}. 

\begin{align}
	E^G = {\int}^{t}_{0} P^G d\tau = \overline{P^G} \times t
	 \label{eq:e_model}
\end{align}

\begin{figure}[ht]
	\centering
	\includegraphics[width=0.42\textwidth]{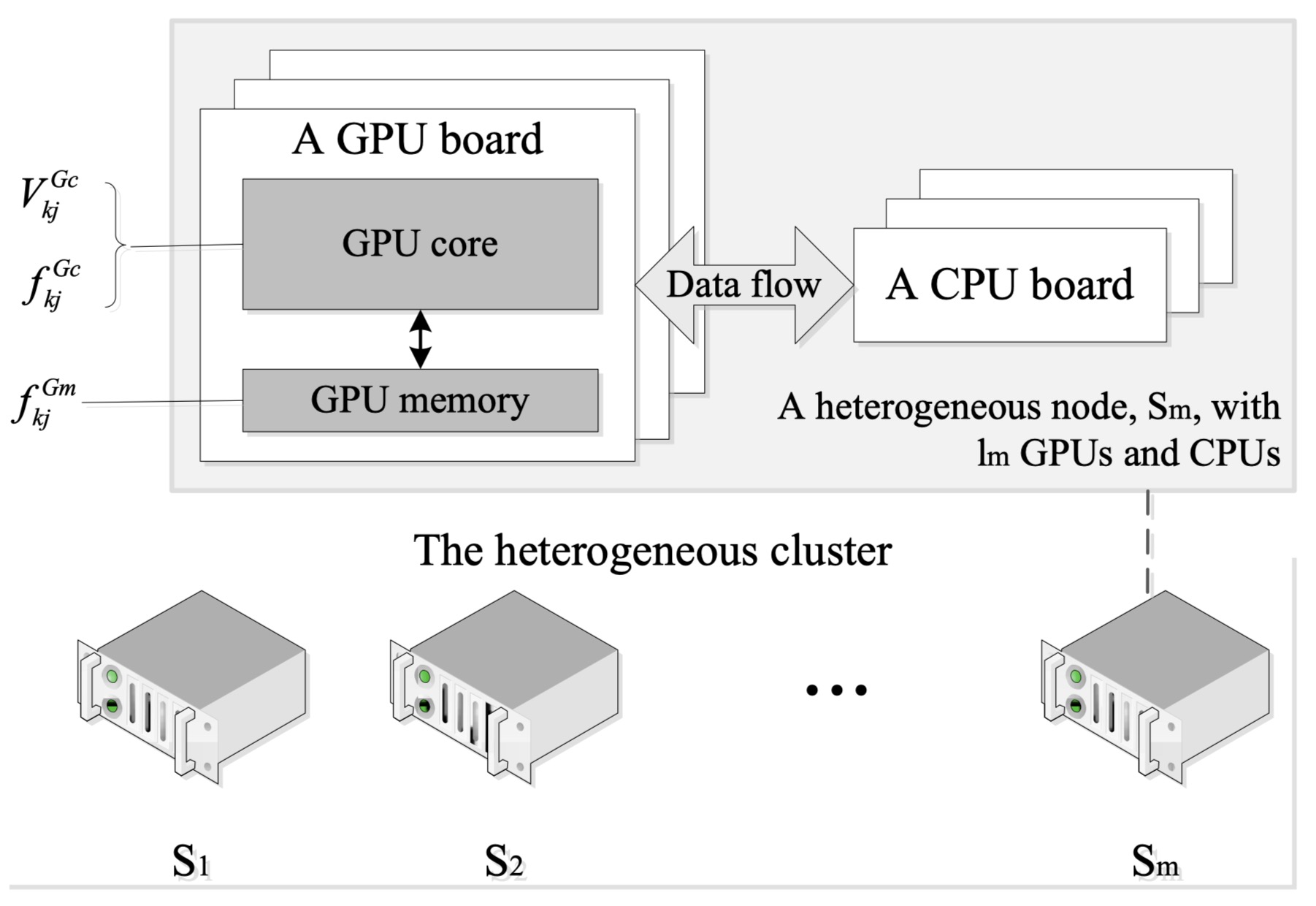}
	\caption{Our studied heterogeneous CPU-GPU cluster with $m$ servers. }\label{fig:cluster}
\end{figure}

\subsubsection{CPU-GPU Cluster Modeling}
Fig. \ref{fig:cluster} shows the model of CPU-GPU cluster considered in this work. In the cluster, there are $m$ servers, each with multiple pairs of CPU-GPU. In this work, we assume that the cluster has only one type of CPU/GPU, but different servers may have different numbers of CPU-GPU pairs. We also assume that each task can be assigned to only one CPU-GPU pair, and one CPU-GPU pair can only execute one task at a moment. In practice, the number of CPU cores in a server should be no less than the number of GPUs. The energy consumption of the additional CPU cores, if any, can be modeled as static energy if they are idle, or be handled by existing CPU energy management techniques if they are used to run CPU jobs.

A CPU-GPU pair can be in one of three states: runtime (or busy), idle, and off. A runtime CPU-GPU pair consumes both dynamic and static power; an idle CPU-GPU pair consumes only low static power; and a turned-off CPU-GPU pair consumes no power. A CPU-GPU pair can be turned off by shutting down the server, which can only happen if there is no job assigned to any of its CPU-GPU pairs. However, there is considerable energy cost from the turning on/off operations. We use $\Delta$ to denote the average energy overhead to turn on/off a single CPU-GPU pair. If any CPU-GPU pair is busy, the other CPU-GPU pairs on the same server without workload have to remain in the idle state.
\begin{align}
	E_J = (P^{G0}+\gamma f^{Gm} + c^G(V^{Gc})^2 f^{Gc}\times& \nonumber \\
	(D(\frac{\delta}{f^{Gc}}+\frac{1-\delta}{f^{Gm}})+t^0)& \label{eq:energy_model}
\end{align}

Since the power consumption of a single CPU core is much less than that of a GPU, it is simplified as a constant in our model, i.e., we include the average CPU runtime power into $P^{G0}$ in Eq. \eqref{eq:power_model} for each GPU-CPU pair. Naturally, the CPU will be kept active if the associated GPU is active, which means that the GPU and CPU share the same execution time to process a task. Under these conditions, the runtime energy consumption ($E_J$) of a CPU-GPU pair to process one single task can be reformulated as Eq. \eqref{eq:energy_model}.
\begin{table}
	\centering
	\caption{Frequently used notations} \label{tab:notation}
	\begin{tabular}{|p{0.4in}|p{2.7in}|} \hline
		Name & Descriptions \\ \hline \hline
		$\textbf{J}$ & The whole task set. \\
		$J_i$ & The $i$-th task in the task set. \\
		$a_i$ & The arrival time of $J_i$. $a_i$ is a unit number. \\
		$d_i$ & The required deadline of $J_i$. We assume $d \leq t^{*} + a_i$. \\
		$\gamma, P^{0}_i$ & Parameters related to the runtime power of $J_i$ \\
		$D_i,t^{0}_i,{\delta}_i$ & Parameters related to the performance of $J_i$ \\
		$\mathscr{P}_i$ & The power consumption function of $J_i$. \\
		$\mathscr{T}_i$ & The performance function of $J_i$. \\
		$P^{*}_i$ & The default runtime power of $J_i$ on one CPU-GPU pair. \\
		$t^{*}_i$ & The default execution time of $J_i$ on one CPU-GPU pair. \\
		\hline
		$S_j$ & The $j$-th server. \\
		$l_j$ & The number of CPU-GPU pairs on $S_j$. \\
		$P^{idle}$ & The idle power consumption of a CPU-GPU pair. \\
		$\Delta$ & The energy overhead of turning on a CPU-GPU pair. \\
		$\rho$ & The threshold for turning off a server. \\
		\hline
		${\kappa}_i$ & The time when the cluster begins executing $J_i$. \\
		${\mu}_i$ & The time when the cluster finishes executing $J_i$. \\
		${\tau}_{kj}$ & The processing time of the $k$-th CPU-GPU pair on the $j$-th server. \\
		$F_j$ & The longest processing time of all the CPU-GPU pairs on the $j$-th server; $F_j = \underset{k}{\text{max}}\{\tau_{kj}\}$. \\
		\hline
		$T$ & The time slot. $T$ is a unit number. \\
		$\textbf{J}(T)$ & The task set arrives at $T$, which has $n(T)$ tasks. \\
		$n(T)$ & The number of tasks in $\textbf{J}(T)$. \\
		$M(T)$ & The number of occupied servers at $T$. $M(T)<m$. \\
		$N^{OFF}$ & The number of offline tasks. $N^{OFF} = n(0)$. \\
		$N^{ON}$ & The number of online tasks. $N^{ON} = \sum_{T\neq 0}n(T)$. \\
		\hline
	\end{tabular}
\end{table}

\subsection{Problem Formulation}
\subsubsection{Assumption}
We formulate our task scheduling problem consider-
ing GPU DVFS as follows. For ease of reference, the major mathematical notations are summarized in Table \ref{tab:notation}. Our CPU-GPU energy optimization problem arises from the
following system setting:

\begin{enumerate}
    \item A CPU-GPU hybrid cluster that consists of $m$ servers, $\textbf{S}={S_1,S_2,...,S_m}$, and the $j$-th server $S_j$, has $l_j$ CPU-GPU pairs;
    \item A task set of n independent and non-preemptive tasks $\textbf{J}={J_1,J_2,...,J_n}$ arriving over time, where the $i$-th task $J_i$ is represented by a tuple $J_i = \{a_i, d_i,\mathscr{P}_i,\mathscr{T}_i\}$, where $a_i$ denotes the arrival time and $d_i$ denotes the task deadline. Every task is non-preemptive that once the processing starts, it cannot be stopped until the task is completed. 
\end{enumerate}

Our objective is to minimize the total energy, $E_{total}$, while satisfying the task deadline constraints:
\begin{align}
    \text{min.}& E^{total} \nonumber \\
    s.t.& {\mu}_i \leq d_i, \forall i. \label{eq:E_min}
\end{align}
where ${\mu}_i$ denotes the time that job $J_i$ finishes. 

Besides, we assume that during the execution of a task, the voltage/frequency setting remains the same. We guarantee the scheduling feasibility by assuming sufficient servers. We make these two assumptions throughout this work.For every $J_i$ in the task set, we need to compute its GPU voltage/frequency configuration as {$V^{Gc}_i, f^{Gc}_i,f^{Gm}_i$}, and its mapping {${\kappa}_i,\phi(J_i)$} where ${\kappa}_i$ denotes the time the cluster begins to execute $J_i$ according to the scheduling algorithm, and $\phi(J_i)$ denotes the assignment of $J_i$. $\phi(J_i)=S^{J_i}_{kj}=1$ indicates that $J_i$ is mapped onto the $k$-th CPU-GPU pair on the $j$-th server. 
\begin{figure}[ht]
	\centering
	\includegraphics[width=0.42\textwidth]{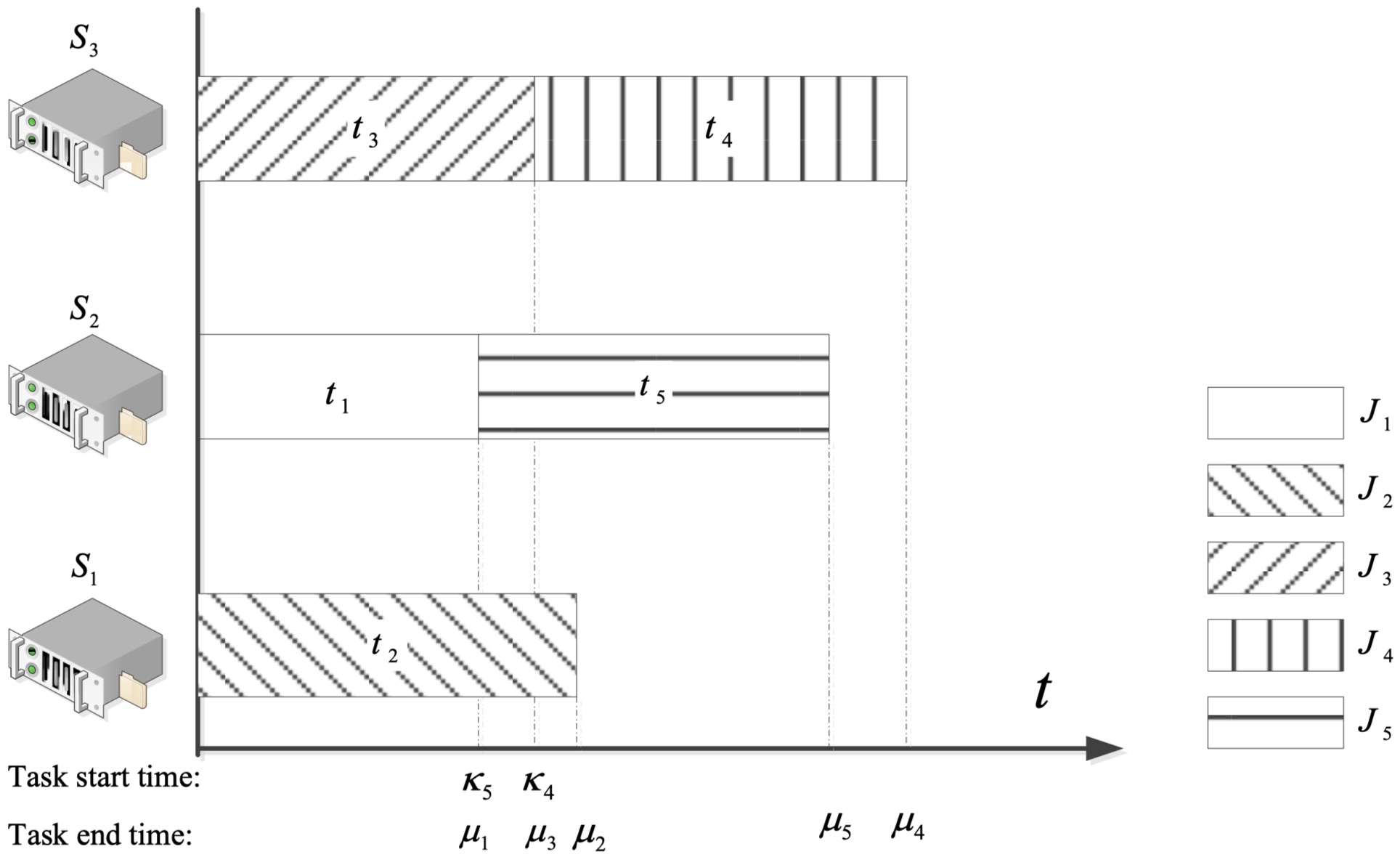}
	\caption{Our studied heterogeneous CPU-GPU cluster with $m$ servers. }\label{fig:sample}
\end{figure}

Fig. \ref{fig:sample} shows an example of our task scheduling solution where every server has just one CPU-GPU pair. There are totally five tasks and three servers. Server $S_1$ is assigned with only one task while $S_2$ and $S_3$ are assigned with two tasks. When the server completes all the assigned tasks, it is turned off and consumes no energy. The total energy consumption equals all the energy the five tasks consume.  

\subsubsection{Offline Case}\label{subsubsec:offline}
For the offline case, all the tasks arrive at $T=0$. $E^{total}$ is shown as Eq. \eqref{eq:E_total_off}. $E^{run}$ represents the runtime energy to process the tasks and $E^{idle}$ represents the total energy consumption by all the CPU-GPU pairs at idle states before the task completion. $F_j$ denotes the longest processing time of all the CPU-GPU pairs on the j-th server. To minimize the total energy, we need to minimize both the runtime energy and the idle energy. 
\begin{align}\label{eq:E_total_off}
    E^{total}&=\underbrace{\sum_{i=1}^{n}P_{J_i}({\mu}_i-{\kappa}_i)}_{E^{run}}+\underbrace{P^{idle}\sum_{j=1}^{M}\sum_{k=1}^{l_j}{(F_j - {\tau}_{kj})}}_{E^{idle}}
\end{align}

\subsubsection{Online Case}
For the online case, all the tasks can arrive at any $T$. The total energy consumption, $E^{total}$, can be decomposed into three parts: $E^{run}, E^{idle},E^{overhead}$, as Eq. \eqref{eq:E_total_on} shows. $E^{run}$ denotes the energy consumption to process all the tasks, i.e., $E^{run}=\sum_{i=1}^{n}E_{J_i}=\sum_{i=1}^{n}P_{J_i}({\mu}_i-{\kappa}_i)$. $E^{idle}$ denotes the idle system energy. It equals the summation of the idle energy of all the CPU-GPU pairs, $E^{idle}=P^{idle}\sum_{j=1}^{m}\sum_{k=1}^{l_j}{\eta}_{kj}$, where ${\eta}_{kj}$ is the total idle period of the $k$-th CPU-GPU pair on the $j$-th server. $E^{overhead}$ denotes the overhead to turn on/off the servers. $E^{overhead} = \omega \Delta$, where $\omega$ is the total number of the turn-on behaviours in the cluster (counted based on the unit of a CPU-GPU pair). $E^{run}$ is closely related to the GPU voltage/frequency setting while the others are more relevant to the scheduling algorithm.
\begin{align}\label{eq:E_total_on}
    E^{total}&=E^{run}+E^{idle}+E^{overhead} \nonumber \\
             &=\sum_{i=1}^{n}P_{J_i}({\mu}_i-{\kappa}_i)+P^{idle}\sum_{j=1}^{m}\sum_{k=1}^{l_j}{\eta}_{kj}+\omega\Delta 
\end{align}

Empirically, when a dynamic turning off mechanism is
involved, $E^{run}$ should be the majority of $E^{total}$. We elaborate in the next section that for a single task, there exists an optimal solution in the DVFS scaling interval to minimize $E^{run}$.

\subsection{Complexity Analysis}
We first prove that the offline task scheduling problem in Section \ref{subsubsec:offline} is NP-hard. Then the online case can be proved similarly. We discuss our problem with two cases: $l_j = 1$ and $l_j \neq 1$. If $l_j = 1$, each server has only one CPU-GPU pair. The servers can be turned off as soon as the tasks are completed so that $E^{idle}$ in Eq. \eqref{eq:E_total_off} is eliminated. Our problem is simplified to find the optimal voltage/frequency setting of every task without missing its deadline. Under this circumstance, the minimized energy is independent of the scheduling algorithm: we can construct a one-to-one task-to-server mapping in the simplest way because of the sufficient server resource, or we can schedule multiple tasks onto one server to reduce the server using. We note the complexity of the optimization problem of each task as $\Phi$, then the complexity of our problem with $l_j = 1$ is $O(n\Phi)$.

The energy efficient task scheduling problem is computational difficult with $l_j \neq 1$. We prove it to be NP-hard by reducing the classical NP problem to it.

\noindent \textbf{Theorem 1.} The task scheduling problem with the objective of minimizing the energy consumption while meeting all task deadlines under the off-line model is NP-hard.

\begin{align}\label{eq:e_total_proof}
    E^{total}&=E^{run}+P^{idle}\sum_{j=1}^{M}F_j-P^{idle}\sum_{j=1}^{M}\sum_{k=1}^{l_j}{\tau}_{kj} \nonumber \\
    &=E^{run}+P^{idle}\sum_{j=1}^{M}F_j-P^{idle}\sum_{j=1}^{n}t_{J_i} \nonumber \\
    &=(E^{run} - P^{idle}\sum_{j=1}^{n}t_{J_i}) + P^{idle}\sum_{j=1}^{M}F_j 
\end{align}

\begin{proof}
We prove the NP hardness with a additional assumption that all the tasks are insensitive to the GPU voltage/frequency scaling, which can be obtained by assuming $P^G = P^{G0}$ in Eq. \eqref{eq:power_model} and $t = t^0$ in Eq. \eqref{eq:time_model}. Then the task processing time and the runtime power consumption of the tasks are fixed, so that $(E^{run} - P^{idle}\sum_{j=1}^{n}t_{J_i})$ in Eq. \eqref{eq:e_total_proof} is a constant. Therefore minimizing $E^{total}$ is equivalent to minimizing $\sum_{j=1}^{M}F_j$.

We consider a special case: m = 1 and a very large common deadline that can always be met. Since the deadline is large, we can also assume $l_1 < n$. Our problem is transformed into assigning $n$ independent tasks to $l_1$ identical CPU-GPU pairs such that the longest processing time (make-span) of all the CPU-GPU pairs is as small as possible, which is identical to the NP-hard multiprocessor scheduling problem \cite{Ibarra1977}. The other way around, the multi-processor scheduling problem is equivalent to our problem when $m = 1$, the task deadlines can always be met and the task length/power consumption does not rely on DVFS.

To summarise, because a simplified version of our problem can be translated into a classical NP problem, our problem is also of NP hardness.
\end{proof}

\section{DVFS Energy Minimization for CPU-GPU Hybrid Clusters}\label{sec:solution}
\subsection{Solution for a Single Task}
As a first step, we consider the following sub-problem:
\textsl{for a single task, given its power and performance models, what is the optimal voltage/frequency setting that minimizes the runtime energy regardless of its deadline?}

Eq. \eqref{eq:argmin_e} shows the mathematical formulation of the problem.
Notice that $V^{Gc}$ and $f^{Gc}$ are correlated variables, and $f^{Gc}$ is upper bounded by a function of $V^{Gc}$, denoted by $g_1(V^{Gc})$.

\begin{align} \label{eq:argmin_e}
    \text{arg min}E_J =&\text{arg min}\{(P^{G0})+\gamma f^{Gm}+c^G(V^{Gc})^2f^{Gc}) \nonumber \\
    &\times (D(\frac{\delta}{f^{Gc}}+\frac{1-\delta}{f^{Gm}})+t^0)\} \nonumber \\
    s.t. V^{Gc}_{min} &\leq V^{Gc} \leq V^{Gc}_{max}, f^{Gm}_{min} \leq f^{Gm} \leq f^{Gm}_{max},\nonumber \\
    f^{Gc}_{min} &\leq V^{Gc} \leq g_1(V^{Gc})
\end{align}

As the memory frequency $f^{Gm}$ is independent of $V^{Gc}$ and $f^{Gc}$, we can analyze core scaling and memory frequency scaling separately. We first consider GPU core voltage and frequency scaling. Given a fixed memory frequency $f^{Gm}_o$ , the solution to Eq. \eqref{eq:argmin_e} satisfies Theorem 1.

\noindent \textbf{Theorem 1.} With a fixed memory frequency, the runtime energy of a task is minimum when the GPU core frequency is maximum corresponding to the GPU core voltage, i.e.,

\begin{align}
    E_{J_{min}}(f^{Gm}_o)=\underset{V^{Gc}}{\text{arg min}}E_J(V^{Gc},g_1(V^{Gc}),f^{Gm}_o). \nonumber
\end{align}

Theorem 1 transforms a two-variable optimization problem into a single-variable optimization problem. It implies that when we scale the GPU core alone to conserve energy, we only need to find an appropriate core voltage and set the core frequency to the largest allowed value.

\begin{figure}[ht]
	\centering
	\includegraphics[width=0.42\textwidth]{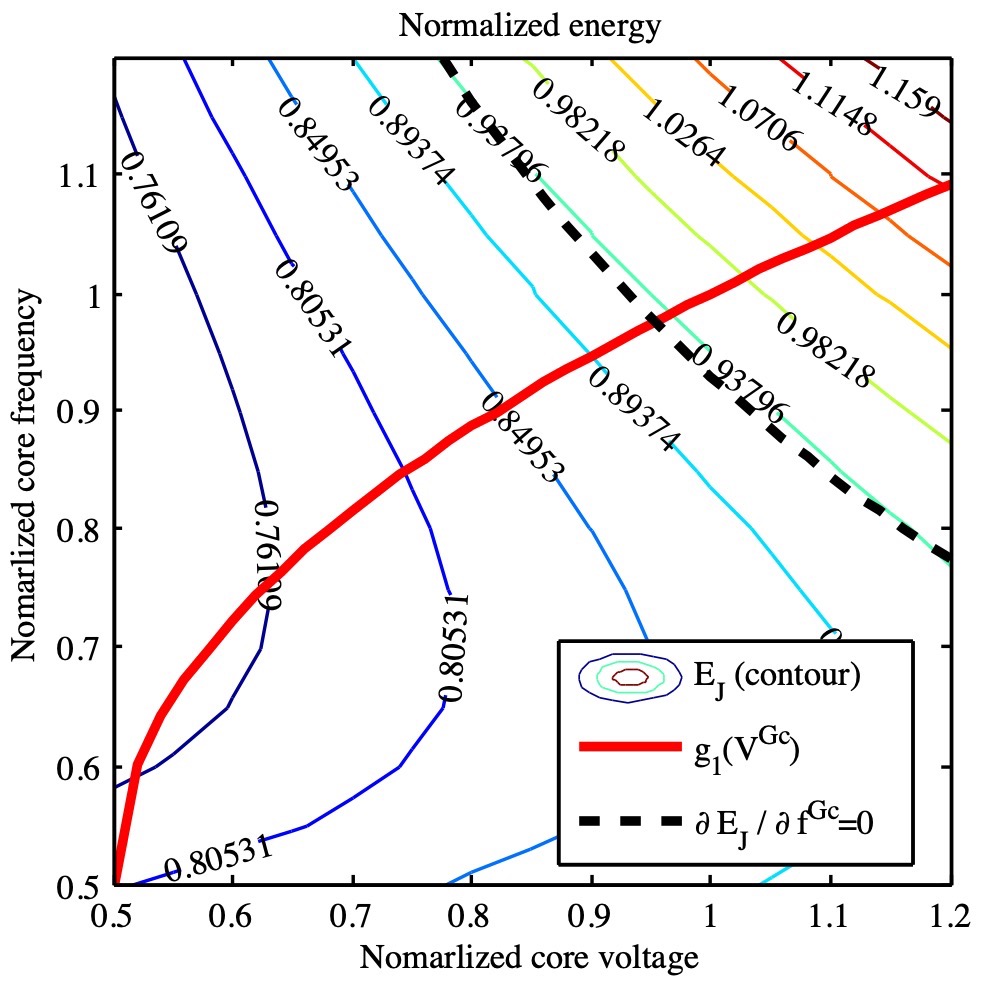}
	\caption{When memory frequency is fixed, the minimum energy depends on the core voltage only. The data is obtained with $P=100+50 f^{Gm} +150{V^{Gc}}^2 f^{Gc}; t=25(0.5/f^{Gc} + 0.5/f^{Gm})+5; g_1(V^{Gc})=\sqrt{(V^{Gc}-0.5)/2}+0.5$ and $f^{Gm}_o=f^{Gm}_\text{max}=1.2$. Note that although we use a specific function for demonstration, the finding holds for other general functions of our GPU DVFS modeling scheme. }\label{fig:func_dvfs}
\end{figure}

\begin{proof}
We obtain the first-order partial derivatives as: $\frac{\partial E_J}{\partial V^{Gc}}=2V^{Gc}c^G f^{Gc}(t^0+D\delta+D(1-\delta)/f^{Gm}_o)$ and $\frac{\partial E_J}{\partial f^{Gc}}=c^G(V^{Gc})^2 (t_0+D(1-\delta)/f^{Gm}_o)-D\delta\frac{(P^0+\gamma f^{Gm}_o)}{(f^{Gc})^2}$. Because $\frac{\partial E_J}{\partial V^{Gc}}>0$, $E_J$ cannot attain its minimum on the interior of the domain, and $E_J$ is a monotonically increasing function of $V^{Gc}$. The minimum is on the boundary of $g_1(V^{Gc})$. $f^{Gc}$ can be eliminated such that finding the minimum of $E_J$ is only related to $V^{Gc}$.

We also give a graphical proof of Theorem 1. In Fig. \ref{fig:func_dvfs}, we plot the contour curves of $E_J,g_1(V^{Gc})$ and $\frac{\partial E_j}{\partial f^{Gc}}=0$ together. As the figure shows, the optimal solution is along the red curve of $g_1(V^{Gc})$, where $g_1(V^{Gc})$ is tangent to the contour curve of $E_J=E_{J_{min}}$.
\end{proof}

We then consider GPU memory frequency scaling alone. If the core voltage and frequency settings are fixed as $V^{Gc}_o$ and $f^{Gc}_o$, we can easily compute the optimal memory frequency by setting $\frac{d E_J}{d f^{Gc}}=0$. We denote $f^{Gm}_{\xi}=\sqrt{(P^0+c{V^{Gc}_o}^2 f^{Gc}_o)D(1-\delta)/(\gamma (t^0+D\delta/f^{Gc}_o))}$, that the optimal memory frequency equals: i)$f^{Gm}_{min}(f^{Gm}_{\xi}<f^{Gm}_{min})$;ii)$f^{Gm}_{\xi}(f^{Gm}_{min}<f^{Gm}_{\xi}<f^{Gm}_{max})$;iii)$f^{Gm}_{max}(f^{Gm}_{\xi} > f^{Gm}_{max})$.

Note that the analytical optimal solution for core scaling or memory scaling alone coincides with that of the practical GPU DVFS experiments. We consider our analytical solution to be quite reasonable. 

Based on the above analysis, the original three-variable problem is transformed into a two-variable optimization problem. Reducing the problem dimension is vital to speeding up the computation.

We now move on to the problem with the task deadline considered: \textsl{what is the optimal voltage/frequency setting for a task without violating the given deadline?}

We denote the previous optimal solution of Eq. \eqref{eq:argmin_e} as {$\hat{f^{Gc}}, \hat{f^{Gm}}, \hat{V^{Gc}}$}. We refer to the execution time obtained by substituting {$\hat{f^{Gc}}, \hat{f^{Gm}}, \hat{V^{Gc}}$} into Eq. \eqref{eq:time_model} as the optimal execution time ($\hat{t}$), and the one without GPU DVFS as the default execution time. The optimal execution time is possibly longer than the default execution time ($t^{*}$).

\noindent \textbf{Definition 1} (Task priority.) We define the task priority according to its optimal execution time. If $d-a < \hat{t}$, the task is deadline-prior; otherwise the task is energy-prior.

Apparently, if a task is deadline-prior, we cannot simply apply the original solution of Eq. \eqref{eq:argmin_e} because scaling down the frequency too much may violate the deadline constraint. For a deadline-prior task, we may need to scale up the voltage/frequency compared to the original optimal setting.

The updated voltage/frequency setting for a deadline-prior task makes the updated execution time ($\hat{t^{'}}$) equal to its allowed time period, i.e., $\hat{t^{'}} = d - a$. We prove that the optimal solution of Eq. \eqref{eq:argmin_e} is on the boundary of the domain. Intuitively, for a deadline-prior task, the additional constraint $f^{Gm} \leq \frac{D(1-\delta)}{d-t^0-\frac{D\delta}{f^{Gc}}}$ shrinks the domain. The updated solution $\hat{f^{Gc}}, \hat{f^{Gm}}, \hat{V^{Gc},\hat{t^{'}}}$ is defined by both $f^{Gc}=\frac{D\delta}{d-t^0-\frac{D(1-\delta)}{f^{Gm}}}$ and $f^{Gc}=g_1(V^{Gc})$. So for the deadline-prior task, $E_{J_{min}}=\underset{f^{Gm}}{\text{arg min}}$, where $t(f^{Gc}_o, f^{Gm})=d-a$ and $f^{Gc}_o=g_1(V^{Gc}_o)$. This is a single-variable optimization problem and can be solved quickly.

\begin{algorithm}
  \caption{Voltage/frequency configuration} \label{algo:vf_conf}
  \begin{algorithmic}[1]
 \renewcommand{\algorithmicrequire}{\textbf{Input:}}
 \renewcommand{\algorithmicensure}{\textbf{Output:}}
    \REQUIRE The task set \textbf{J} with the property table.
    \ENSURE $n_1$, \textbf{J} with the optimized voltage/frequency. 
    \STATE $n_1 \leftarrow 0$; //$n_1$: the number of deadline-prior tasks
    \FOR{$i$ = 1 to $n$}
        \STATE compute \{$\hat{V^{Gc}_i}, \hat{f^{Gc}_i}, \hat{f^{Gm}_i}, \hat{t_i}$\};
        \IF{$\hat{t_i} > d_i$}
              \STATE // $J_i$ is deadline-prior, update the configuration
              \STATE compute \{$\hat{V^{Gc}_i}^{'}, \hat{f^{Gc}_i}^{'}, \hat{f^{Gm}_i}^{'}, \hat{t_i}^{'}$\};
              \STATE $n_1 \leftarrow n_1 + 1$;
        \ENDIF
    \ENDFOR
  \end{algorithmic}
\end{algorithm}

\subsection{Solution for Multiple Tasks}
If we apply the derived optimal solution for each deadline-prior and energy-prior task in the task set, we obtain a fixed computed task length $\frac{\hat{t}}{\hat{t^{'}}}$, and a minimized $E^{run}$. In this
section, we consider the problem: \textsl{given the optimal computed task length of each task, how to schedule a number of tasks on the CPU-GPU cluster?} In the following, we first propose our solution, the EDL $\theta$-readjustment scheduling algorithm, for the offline case. Then we extend the EDL $\theta$-readjustment scheduling algorithm to the online problem, where we combine the GPU DVFS and dynamic resource sleep (DRS).

After executing Algorithm \ref{algo:vf_conf}, we get the number of the deadline-prior tasks in the task set ($n_1$) and the task length of each task ($\hat{t_i}$). Apparently, we need to assign the deadline-prior tasks with appropriate configuration to the cluster as earlier as possible, otherwise the task deadlines would be missed. The deadline-prior tasks would occupy $n_1$ CPU-GPU pairs.

For the remaining energy-prior tasks, we sort them in the deadline-increasing order and assign them to the CPU-GPU pairs sequentially, which is named as the earliest-deadline-first (EDF) scheduling. The EDF algorithm is proved to be optimal in terms of feasibility \cite{Liu1973}.

As a typical strategy to handle online job scheduling, we divide time into equal time slots, and schedule newly arrived tasks in a time slot as a batch. The duration of a time slot should be significantly shorter than the average job execution time. The system is initiated with a set of offline tasks, which arrive at time slot $T = 0$; and the online tasks arrive at different time slots $T \neq 0$. The set of tasks arriving at time slot $T$ is denoted by $\textbf{J}(T)$. At the beginning of each time slot, we sort the newly arrived tasks in deadline-increasing order and assign them sequentially. This is referred to as earliest-deadline-first (EDF) scheduling, which is proved to be optimal in terms of feasibility \cite{Liu1973}.

The task mapping follows a simple principle that always tries to assign the task with the derived optimal task length to the CPU-GPU pair with the lightest workload. The objective is mainly minimizing $E^{run}$. We define another parameter, $\theta$, to strike a better balance between the two conflicting objectives: minimizing $E^{run}$ and minimizing $E^{idle}$.

\noindent \textbf{Definition 2} (Task deferral threshold). Given $\hat{t^{'}_i}$ as the optimal execution time with minimized runtime energy of $J_i$, instead of fixing the task execution time as $\hat{t^{'}}$, we allow it to vary in the interval of $[\theta \hat{t^{'}_i}, \hat{t^{'}_i}]$, $0 < \theta \leq 1$ by readjusting the frequency setting, in order to further reduce the total energy.

$\theta$ describes how much we can sacrifice the runtime energy for a shorter make-span and less occupied servers. It applies proper voltage/frequency readjustments during the process of task scheduling. When a $\theta$-readjustment is applied, we allow the non-optimal voltage/frequency setting for the energy-prior task in order to make usage of the currently alive servers with idle CPU-GPU pair(s). This behavior transfers a number of energy-prior tasks into deadline-prior tasks. By default, $\theta = 1$ and no readjustment is allowed. By varying the value of $\theta$, we actually control the maximum allowed portion of such transformation we can make in a task set. Because $\theta$ is designed to further reduce the idle energy, intuitively $\theta < 1$ is effective only when $l > 1$ and the idle energy is non-negligible.

We demonstrate some examples about the effectiveness of $\theta$ parameter with an artificial test set listed in Table \ref{tab:ex_theta}. The five tasks have the same default execution time as $t=5+25(\frac{\delta}{f^{Gc}}+\frac{1-\delta}{f^{Gm}})$. The idle power consumption of a CPU-GPU pair is 30 W. Every two CPU-GPU pairs are grouped into one server. We assume $\theta = 0.9$ and $\gamma = 0$. After executing Algorithm 1, we get the optimal execution time of each task, as the last column in Table 5.2 shows. Of the five tasks, $J_2$ is deadline-prior and need to be assigned as soon as possible. We note the CPU-GPU pairs as $S_{11}, S_{12}, S_{13},...,$ etc. Firstly we assign $J_2$ to $S_{11}$. Then we sort other tasks according to EDF: {$J_1, J_3, J_4, J_5$} and decide the scheduling for them one by one. $J_1$ cannot be assigned to $S_{11}$ so that we schedule it to $S_{12}$. In the second step, because $S_{12}$ has the shorter processing time, we try to assign $J_3$ to $S_{12}$. The remaining time before the deadline of $J_3$ is $d_3 - \hat{t}$ = 34.17, which is shorter than $\hat{t_3}$. We then consider the $\theta$ readjustment with an allowable interval of $t_3 \in [31.90, 35.44]$. The minimum execution time of $J_3$ is 25.83. The remaining time fits into the interval so that we can re-adjust the runtime voltage/frequency to make the execution time equal 34.17. Likewise, we assign $J_4$ to $S_{11}$ and $J_5$ to $S_{12}$ to have the final mapping: $S_{11}(J_2, J_4)$, $S_{12}(J_1, J_3, J_5)$. With the same task set but $\theta$ = 1, we would get another mapping: $S_{11}(J_2)$, $S_{12}(J_1, J_4)$, $S_{21}(J_3, J_5)$. The latter one would consume more energy. $\theta$ = 0.9 is more effective in saving the system energy.

\begin{table}\label{tab:ex_theta}
	\centering
	\caption{An example of a task property table}
	\begin{tabular}{|c||cc|cc||cc||cc|} \hline
		Task & $P^0$ & $P^*$ & $t^0$ & $t^*$ & $\delta$ & $d$ & $\hat{P}$ & $\hat{t}$ \\ \hline 
		$J_1$ & 100 & 300 & 5 & 30 & 0 & 50 & 125.23 & 25.83 \\
		$J_2$ & 100 & 300 & 5 & 30 & 1.0 & 36 & 176.31 & 36 \\
		$J_3$ & 100 & 300 & 5 & 30 & 0.5 & 60 & 135.20 & 35.44 \\
		$J_4$ & 100 & 300 & 5 & 30 & 0.8 & 100 & 141.39 & 39.10 \\
		$J_5$ & 100 & 300 & 5 & 30 & 0.2 & 300 & 127.60 & 30.86 \\
		\hline
	\end{tabular}
\end{table}

\subsubsection{Offline Scheduling}
We describe our EDL $\theta$-readjustment scheduling algorithm in Algorithm \ref{algo:edl_off}. Line 16 considers the relationship between $\theta \hat{t}$ and $t_{min}$, where $t_{min}$ denotes the minimum execution time of a task, due to the natural fact that an execution time shorter than $t_{min}$ is unreachable. When the $\theta$ readjustment is effective, as listed in lines 17-19, we need to configure the voltage/frequency setting again. The problem is similar to the deadline-constrained optimization where $d_r - {\mu}_{SPT}$ can be regarded as the new task deadline.
\begin{algorithm}
	\caption{The EDL $\theta$-readjustment scheduling algorithm} \label{algo:edl_off}
	\begin{algorithmic}[1]
		\renewcommand{\algorithmicrequire}{\textbf{Input:}}
		\renewcommand{\algorithmicensure}{\textbf{Output:}}
		\REQUIRE $n_1$, \textbf{J} with the optimized voltage/frequency, $\theta$.
		\ENSURE $m_1, \{f_{i}^{Gc}, V_{i}^{Gc}, f_{i}^{Gm}, {\kappa}_i, {\mu}_i\}$ and the mapping of $J_i, m_1$ CPU-GPU pairs with workloads.
		\FORALL{deadline-prior tasks}
		\STATE Schedule each of them to a CPU-GPU pair;
		\ENDFOR
		\STATE $n_2 \leftarrow n - n_1$; // $n_2$: the number of energy-prior tasks
		\FORALL{energy-prior tasks}
		\STATE $\{J_1,...,J_r,...,J_{n_2}\} \leftarrow$ sort them in EDF order;
		\ENDFOR
		\STATE $m_1 \leftarrow n_1$; 
		\STATE // $m_1$: the number of occupied CPU-GPU pairs
		\FOR{$r$ = 1 to $n_2$}
		\STATE ${\mu}_{SPT} \leftarrow $ min$\{{\mu}_1,...,{\mu}_{m_1}\}$; 
		\STATE // Find the CPU-GPU pair, $S_{SPT}$, with the shortest processing time
		\IF{$d_r - {\mu}_{SPT} \ge \hat{t_r}$}
		\STATE Assign $J_r$ to $S_{SPT}$;
		\ELSE
		\STATE $t_{\theta} \leftarrow$ max$\{\theta \hat{t_r}, {t_r}_{\text{min}}\}$; \COMMENT{${t_r}_{\text{min}}$: the minimum execution of $J_r$}
		\IF{$d_r - {\mu}_{SPT} \ge t_{\theta}$}
		\STATE \{$\hat{V_{i}^{Gc}}^{'}, \hat{f_{i}^{Gc}}^{'}, \hat{f_{i}^{Gm}}^{'}$\} $\leftarrow \hat{t_i}^{'}=d_r - {\mu}_{SPT}$; \COMMENT{$\theta$-readjustment DVFS is allowed for $J_r$, reconfigure $J_r$}
		\STATE Assign $J_r$ to $S_{SPT}$;
		\ELSE
		\STATE Assign $J_r$ to a new CPU-GPU pair;
		\STATE $m_1 \leftarrow m_1 + 1$;
		\ENDIF
		\ENDIF
		\ENDFOR
	\end{algorithmic}
\end{algorithm}
By executing Algorithm \ref{algo:vf_conf} and Algorithm \ref{algo:edl_off}, we are able to assign $n$ tasks to $m_1$ CPU-GPU pairs. In the last step, we execute Algorithm \ref{algo:server_map} to group $m_1$ identical CPU-GPU pairs into servers. We sort the CPU-GPU pairs based on their execution time and cluster them accordingly. This grouping method ends in the minimum total idle time period.

To summarize, we solve the problem by dividing the solution into three phases. In the first phase, we compute an optimal voltage/frequency setting for each task so that $E^{run}$ in Eq. \eqref{eq:E_total_off} is minimized. In the second phase, we pack the tasks to different servers of which the task lengths are decided by the optimal setting. We also introduce the $\theta$ parameter to discard the minimum $E^{run}$ for reducing $E^{idle}$ when appropriately. Finally we cluster the CPU-GPU pairs to get the final task-to-server scheduling scheme and the total energy consumption.

\begin{algorithm}
  \caption{Server mapping} \label{algo:server_map}
  \begin{algorithmic}[1]
 \renewcommand{\algorithmicrequire}{\textbf{Input:}}
 \renewcommand{\algorithmicensure}{\textbf{Output:}}
    \REQUIRE $m_1$, the task mapping.
    \ENSURE $E^{total}, S_{kj}^{J_i}$ of $J_i$.
    \FORALL{$m_1$ occupied CPU-GPU pairs}
        \STATE Sort in $\mu$-descending order;
        \STATE $M, E^{total} \leftarrow$ Group them into $M$ separate servers according to $l_j$;
    \ENDFOR
  \end{algorithmic}
\end{algorithm}

\subsubsection{Online Scheduling}
In the following, we extend Algorithm \ref{algo:edl_off} to the online problem. Algorithm \ref{algo:edl_framework} shows our online EDL scheduling framework. At $T = 0$, we process the initial set of tasks. Line 1 would output $M(0)$ occupied servers, and the task mapping solution for all the initial tasks.
\begin{algorithm}
	\caption{Online EDL scheduling framework} \label{algo:edl_framework}
	\begin{algorithmic}[1]
		\renewcommand{\algorithmicrequire}{\textbf{Input:}}
		\renewcommand{\algorithmicensure}{\textbf{Output:}}
		\REQUIRE \textbf{J}, \textbf{S}, $\theta$.
		\ENSURE $M(t)$, the corresponding runtime power state of the $M(t)$ occupied servers, \{$f_{i}^{Gc}, V_{i}^{Gc}, f_{i}^{Gm}, {\kappa}_i, {\mu}_i$\} and the mapping of $J_i, \forall i$.
		\FORALL{$T > 0$}
		\STATE Process the tasks leaving at the current time slot;
		\STATE Turn off the idle servers when appropriate;
		\IF{there are arriving tasks}
		\STATE Assign the tasks to the server according to Algorithm \ref{algo:edl_on}, and turn on the servers if needed;
		\ENDIF
		\ENDFOR
	\end{algorithmic}
\end{algorithm}

Our online scheduling has three major components: processing leaving tasks, turning off the servers, and assigning the newly arrived tasks. We describe these components one by one as follows.

\textsl{Processing leaving tasks.} At each time slot, we identify the set of tasks with $\ceil{{\mu}_i} = T$. We set the corresponding CPU-GPU pairs to idle during the time period of $({\mu}_i, T)$. If a CPU- GPU pair still has tasks to process, we assign the next task to it at time slot $T$.

\textsl{Turning off the servers.} After processing the departured tasks, we dynamically turn off the servers using the DRS technique. We do not turn off the server immediately when there is no task to execute on it in the next time slot. Instead, we turn it off after all the CPU-GPU pairs on this server have been idle for at least a period of $\rho$. This strategy avoids frequent turn-on energy overhead in the case of job arrivals in the near future, at the price of slightly increased idle energy consumption.

\textsl{Assigning the newly arrived tasks.} Algorithm \ref{algo:edl_on} shows our assignment strategy for task set $\textbf{J}(T)$. We divide the solution into two phases. In the first phase (lines 1-4), we compute the optimal voltage/frequency setting that minimizes the runtime energy for each task. With this setting, we obtain a fixed task length of each task. Then in the second phase (lines 5-23), we pack the tasks to servers according to the obtained task lengths and the task deadlines. We always try to assign a task to the CPU-GPU pair with the lightest workload (lines 6-9). Note that in line 6 we need to find the larger value of ${\mu}_{SPT}$ and $T$, in the case that the CPU-GPU pair has been idle. If the task cannot fit into the selected pair, we check if a voltage/frequency readjustment is possible by setting its task length equal to the remaining time before the deadline (line 14). In line 18, if the task cannot fit into any active CPU-GPU pairs even with the readjustment, we assign it to a new CPU-GPU pair. We turn on the server containing this CPU-GPU pair and set its other CPU-GPU pairs to the idle state.

The complexity of this algorithm is $n(\text{log} n+\Phi+m)$, where $\Phi$ denotes the complexity of solving the optimization problem in the previous section.

\begin{algorithm}
	\caption{The EDL $\theta$-readjustment upon task arrival} \label{algo:edl_on}
	\begin{algorithmic}[1]
		\renewcommand{\algorithmicrequire}{\textbf{Input:}}
		\renewcommand{\algorithmicensure}{\textbf{Output:}}
		\REQUIRE $T, M^{'}(T), \textbf{J}(T), n(T), \theta, l$.
		\ENSURE the voltage/frequency setting and the mapping of $\textbf{J}(T), M(T)$. // $M^{'}(T)$: the number of occupied servers after turning off the servers in the previous part
		\FORALL{tasks in $J$(T)}
		\STATE Find the optimal voltage/frequency setting without missing the deadline for each task;
		\STATE \{$J_1,...,J_r,...,J_{n(T)}$\} $\leftarrow$ sort the tasks according to the computed optimal length in EDF order;
		\ENDFOR
		\FOR{$r=1$ to $n(T)$}
		\STATE ${\mu}_{SPT} \leftarrow $ min$\{{\mu}_1,...,{\mu}_{M^{'}(T)*l}\}$; 
		\STATE // Find the CPU-GPU pair, $S_{SPT}$, with the shortest processing time
		\IF{$d_r - {\mu}_{SPT} \ge \hat{t_r}$}
		\STATE Assign $J_r$ to $S_{SPT}$;
		\ELSE
		\STATE $t_{\theta} \leftarrow$ max$\{\theta \hat{t_r}, {t_r}_{\text{min}}\}$; // ${t_r}_{\text{min}}$: the minimum execution of $J_r$
		\IF{$d_r - \text{max}(T, {\mu}_{SPT}) \ge t_{\theta}$}
		\STATE \{$\hat{V_{i}^{Gc}}^{'}, \hat{f_{i}^{Gc}}^{'}, \hat{f_{i}^{Gm}}^{'}$\} $\leftarrow \hat{t_i}^{'}=d_r - \text{max}(T, {\mu}_{SPT})$; // $\theta$-readjustment DVFS is allowed for $J_r$, reconfigure $J_r$
		\STATE Assign $J_r$ to $S_{SPT}$;
		\ELSE
		\STATE Assign $J_r$ to a new CPU-GPU pair;
		\STATE Set the other CPU-GPU pairs on this server to idle;
		\STATE $M^{'}(T) \leftarrow M^{'}(T) + 1$;
		\ENDIF
		\ENDIF
		\ENDFOR
		\STATE $M(T) \leftarrow M^{'}(T)$;
	\end{algorithmic}
\end{algorithm}

\begin{algorithm}
	\caption{The bin-packing scheduling algorithm} \label{algo:bin_packing}
	\begin{algorithmic}[1]
		\renewcommand{\algorithmicrequire}{\textbf{Input:}}
		\renewcommand{\algorithmicensure}{\textbf{Output:}}
		\REQUIRE $\textbf{J}, \textbf{S}$.
		\ENSURE $M(t)$, the voltage/frequency setting and the mapping of $\textbf{J}$.
		\FORALL{offline tasks}
		\STATE \{$J_1,...,J_r,...,J_{N^{OFF}}$\} $\leftarrow$ sort the offline tasks in the earliest-deadline-first order;
		\ENDFOR
		\FOR{$r=1$ to $N^{OFF}$}
		\STATE Compute the optimal \{$\hat{V_{r}^{Gc}}, \hat{f_{r}^{Gc}}, \hat{f_{r}^{Gm}}, \hat{t_{r}}$\} for $J_r$, and the optimal task utilization $\hat{u_r}$;
		\STATE Assign $J_r$ to the CPU-GPU pairs according to the worst-fit heuristic, where the utilization of a CPU-GPU pair is no larger than 1 \cite{Liu1973};
		\ENDFOR
		\FORALL{$T > 0$}
		\STATE Processing the tasks leaving at the current time slot;
		\STATE Turn off the idle servers when appropriate;
		\IF{$\textbf{J}(T) \neq \emptyset$}
		\STATE Sort $\textbf{J}(T)$ in EDF order;
		\FOR{$r=1$ to $n(T)$}
		\STATE Compute the optimal  \{$\hat{V_{r}^{Gc}}, \hat{f_{r}^{Gc}}, \hat{f_{r}^{Gm}}, \hat{t_{r}}$\} for $J_r$;
		\STATE Assign $J_r$ to the CPU-GPU pairs according to the first-fit heuristic, following the criteria in \cite{Liu2012}, and turn on the servers when needed;
		\ENDFOR
		\ENDIF
		\ENDFOR
	\end{algorithmic}
\end{algorithm}
\section{Performance Evaluation} \label{sec:experiment}

In this section, we evaluate the performance of our EDL scheduling algorithm, including the offline case (i.e. a batch of jobs arrive at $T=0$) and the online case (i.e. jobs arrive randomly at any $T>0$). We first describe our simulation configuration in Section \ref{subsec:sc}, including GPU scaling configuration, cluster configuration and task set generator. We apply the same GPU and cluster configuration for both offline and online cases. As for task set generation, we adopt different task set characteristics for the offline and online cases. Then we present the experimental results and simulation results of GPU DVFS on a single task in Section \ref{subsec:dvfs_single}. After that, we report the experimental results of our EDL scheduling algorithm on the offline and online cases in Section \ref{subsec:ex_edl_off} and \ref{subsec:ex_edl_on}, respectively.

\subsection{Simulation Configuration}\label{subsec:sc}
\subsubsection{The GPU Scaling Interval} 
In order to assess the effectiveness of GPU DVFS, we first conduct real experiments by a commercial power meter to measure the real power consumption and record the execution time for a set of benchmark applications under different DVFS settings on an Nvidia Pascal GPU, GTX 1080Ti \cite{wang2020dvfs}. We then conduct simulations based on the gathered data sets. Literally the range of scalable GPU voltage and frequency varies among different GPU products. Without loss of generality, we compute the normalized values of $f^{Gc}, V^{Gc}$ and $f^{Gm}$ based on the factory default values, instead of the absolute values.

In our experiments, for each fixed $V^{Gc}$, we gradually scale up $f^{Gc}$ until the GPU board becomes unstable to get the corresponding $f^{Gc}_{max}$. We fit the $f^{Gc}_{max} = g_1(V^{Gc})$ relationship according to the measurement data as $g_1(x)=\sqrt{(x-0.5)/2}+0.5$. On our real GPU platform, the scaling interval is: $V^{Gc} \in
[0.8, 1.24], f^{Gc}\in[0.89, g_1(V^{Gc})], f^{Gm} \in [0.8, 1.1]$. However, we consider this work from a perspective of discussing the potential of DVFS, thus in the simulation we allow a wider analytically scaling interval to be: $f^{Gm} \in [0.5, 1.2], V^{Gc} \in [0.5, 1.2]$, and $f^{Gc}(V^{Gc}) \in [0.5, g_1(V^{Gc})]$ where $f^{Gc}_{max} \approx 1.09$. The GPU voltage/frequency in the interval is continuously adjustable. In this analytical interval, the power consumption $\mathscr{P}$ is strictly convex.

\subsubsection{Cluster Configuration}
On our real CPU-GPU platform, it has $P^{idle} = 37$ W (24 W for the CPU and 13 W for the GPU), and $(V^{Gc}, f^{Gc}, f^{Gm}) = (1, 1, 1)$ indicating the corresponding GPU configuration of (1.05 V, 1800 MHz, 5000 MHz). We use the data for each simulated CPU-GPU pair.

We choose $\rho=\floor{\Delta/P^{idle}}$, which is derived from the case that the task arriving at the next time slot would occupy the same server and each server has a single CPU-GPU pair ($P^{idle}\rho \leq \Delta$). We set $\Delta=90$ Watts and $P^{idle} = 37$ Watts to have $\rho=2$. Note that there might be other substitutions for $\rho$ which provides better energy conserving performance, but since this paper focuses on DVFS technique, we stay with a simple strategy in the setting of $\rho$.

In addition, we assume there are at most 2048 CPU-GPU pairs, and every 1/2/4/8/16 CPU-GPU pairs are grouped into a server, i.e., $\sum_{j=1}^{m}l_j=2048, l_j=1/2/4/8/16$.

\subsubsection{Task Set Generator}
The simulated task property is also based on real data. We measure the average runtime power and the execution time with 5 $V^{Gc}/f^{Gc}$ samples and 4 $f^{Gm}$ samples, of 20 GPU benchmark applications in \cite{wang2020dvfs}. For each task, we use $5\times4=20$ data samples to fit the power consumption to get $\{P_{i}^0,{\gamma}_i,P_{i}^*\}$, and the performance to obtain $\{t_{i}^0,t_{i}^*,{\delta}_i\}$. We utilize these 20 applications with good fitting results to construct an application library, where $P_{i}^*\in[175, 206], {\gamma}_i / P_{i}^*\in[0.1, 0.2], P_{i}^0 / P_{i}^*\in[0.20, 0.41], {\delta}_i\in[0.07, 0.91], D\in[1.66, 7.61]$ and $t_{i}^0\in[0.1, 0.95]$.

As for the offline cases, we quantize the workload of the task set by the task set utilization ($U_{\textbf{J}}$), which is defined as the summation of the task utilization (i.e., $U_{\textbf{J}}=\frac{\sum_{i=1}^n u_i}{\overline{u_i}\sum_{j=1}^m l_j}$.) based on 1024 CPU-GPU pairs. For example, $U_{\textbf{J}}=1$ represents that the summation of the utilization of all the tasks is 1024. In our simulations, we use 2048 pairs as the baseline rather than 1024, for ease of discussing the mapping feasibility. For example, if a task set with $U_{\textbf{J}}=1$ occupies more than 1024 CPU-GPU pairs, then there is no feasible solution for this task set with the corresponding scheduling algorithm. Since we allow a maximum of 2048 CPU-GPU pairs, we can still compute the actual occupied processor numbers. A general bound to guarantee the feasibility is $U_{\textbf{J}} \le \frac{ml}{2ml-1}$ \cite{lopez2004utilization}.

As for the online cases, the simulated task properties are also based on real measurement results. We still use the task set utilization to quantize the offline tasks and the online tasks. We assign the initial offline task set utilization and the online task set utilization as $U_{\textbf{J}^{OFF}}$ and $U_{\textbf{J}^{ON}}$ separately, which contain $N^{OFF}$ and $N^{ON}$ tasks. In this work, $U_{\textbf{J}^{OFF}}= 0.4$ and $U_{\textbf{J}^{ON}} = 1.6$. We simulate the task arrival in one day and choose the basic time unit as one minute, i.e., $T \in [1, 1440]$. We generate the number of arriving tasks at each time slot, $n(T), T \in [1, 1440]$ according to the Poisson distribution and refine it until $\sum_{T=1}^{1440} n(T) = N^{ON}$. At each time slot, we pick the $(\sum_{o=1}^{T-1}n(T)+1)$-th to $(\sum_{o=1}^{T}n(T))$-th task from the online task set to construct the current arrival tasks, of which $a_i = T$.

Unless the total task utilization is greater than $U_{\textbf{J}}$, at each time, we randomly pick out a task from the library and generate an integer in [10, 50]. We multiply the integer with $\{t^0,t^*\}$ of the task to generate tasks of various lengths. We then generate the task utilization according to the uniform distribution in (0, 1), thus the expectation of the task utilization is 0.5, i.e, $\overline{u_i}=0.5$. We use the default execution time and the task utilization to derive the task deadline, i.e., $d_i=a_i+t^{*}_i/u_i$. We modify the property of the last task to make the total task utilization exactly equal to $U_{\textbf{J}}$. With the above methods, we are able to generate a task set of $n$ tasks, with a total utilization of $U_{\textbf{J}}$.

For each $l$ and $\theta$ setting, we generate 100 groups of the above task sets. We compute the average $E^{idle}$, $E^{overhead}$, $E^{run}$ and the number of occupied servers separately. The total energy consumption is a summation of $E^{idle}$, $E^{overhead}$ and $E^{run}$.

\subsection{DVFS Effect on a Single Task} \label{subsec:dvfs_single}
\begin{figure}[htbp]
	\centering     %%% not \center
	\includegraphics[width=0.92\linewidth]{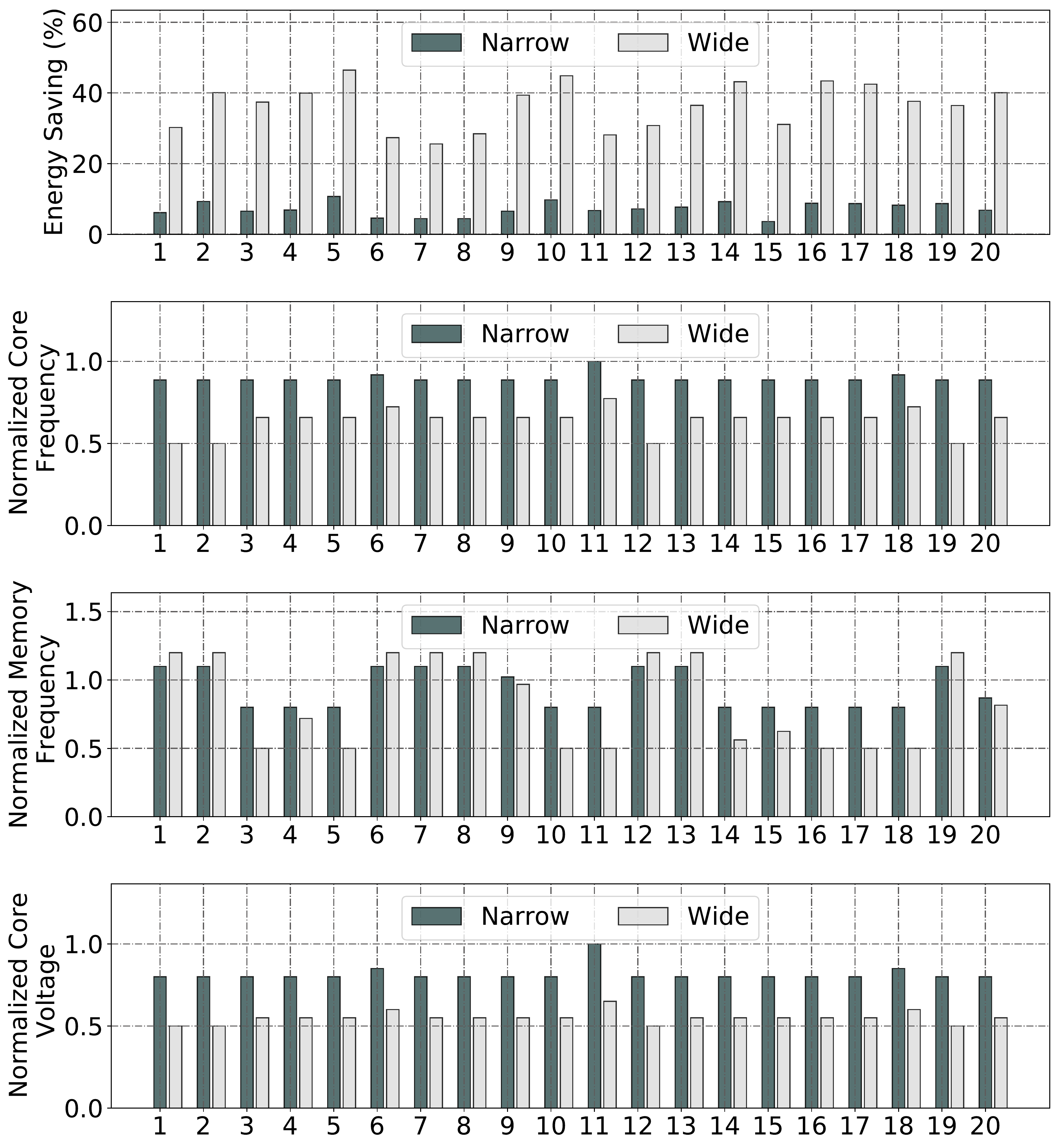}
	\label{fig:single_exp}
	\caption{The energy consumption and the optimal voltage/frequency setting of the 20 benchmark applications. The $x$-axis stands for the application index}\label{fig:single_dvfs}
\end{figure}

In this subsection, we present the experimental results and simulation results of GPU DVFS on a single task. Our realistic experiments show that the average energy conservation of 20 benchmarks is 4.3\% for GTX 1080Ti, which is close to the results in \cite{wang2020dvfs}. The reason for this low effect is (1) The static power $P^{G0}$ takes a big portion in the total power consumption; (2) The scaling intervals of $f^{Gc}$ and $f^{Gm}$ are narrow. According to our measurements, scaling down the core voltage and applying the corresponding maximum allowed core frequency can significantly reduce the energy consumption. The memory frequency scaling influences the energy consumption mostly on the execution time, and different applications have different optimal memory frequency settings. Thus, to discuss the potential of GPU DVFS, we shrink the static power $P^{G0}$ and enlarge the scaling intervals of two frequencies in our simulations.  Fig. \ref{fig:single_exp} shows the derived optimal voltage/frequency setting and the corresponding energy saving of our 20 benchmark applications. Legend 'Wide' stands for our simulated scaling interval, and 'Narrow' stands for the realistic scaling interval. For both intervals, the optimal core voltage/frequency is relatively low, close to the allowed lowest setting. The optimal memory frequency varies, depending on the application characteristics. The derived optimal voltage/frequency settings coincide with our measurements. The energy conservation of the 'Wide' case finally achieves an average value of 36.4\%.

\subsection{Offline EDL DVFS Performance} \label{subsec:ex_edl_off}

In this section, we evaluate the performance of our EDL scheduling algorithm on the offline case. We compare it to the typical EDF-BF/WF algorithm \cite{aydin2003} and the LPT-FF algorithm \cite{Liu2012}. We modify the typical algorithms to fit our model. To make the other three algorithm support DVFS, in the first step, we apply the voltage/frequency scaling according to Algorithm \ref{algo:vf_conf} for each task, so each task would get an optimized task length. In the second step, we firstly schedule those deadline-prior tasks to the servers, and then schedule the energy-prior tasks according to the EDF-BF, EDF-WF, and LPT-FF algorithms. Ultimately, we compute the overall energy consumption and the occupied servers according to Algorithm \ref{algo:server_map}. The overall workflow is similar to our EDL algorithm. 

We conduct the scheduling algorithms first without GPU DVFS and then with it. We refer to the energy consumption without DVFS and $l=1$ ($E^{idle}=0$) as the baseline. We compare the energy consumption with DVFS to the baseline energy and compute the energy saving. We also evaluate the number of occupied servers. The larger energy saving and the fewer servers indicate better algorithm performance. For the figures in this section, legend EDF-SPT denotes our EDL algorithm.

\subsubsection{Baseline Performance}
We first investigate the performance of the four scheduling algorithms without DVFS. We perform our EDL algorithm with $\theta=1$.

\begin{figure}[htbp]
	\centering     %%% not \center
	\subfloat[]
	{
		\includegraphics[width=0.455\linewidth]{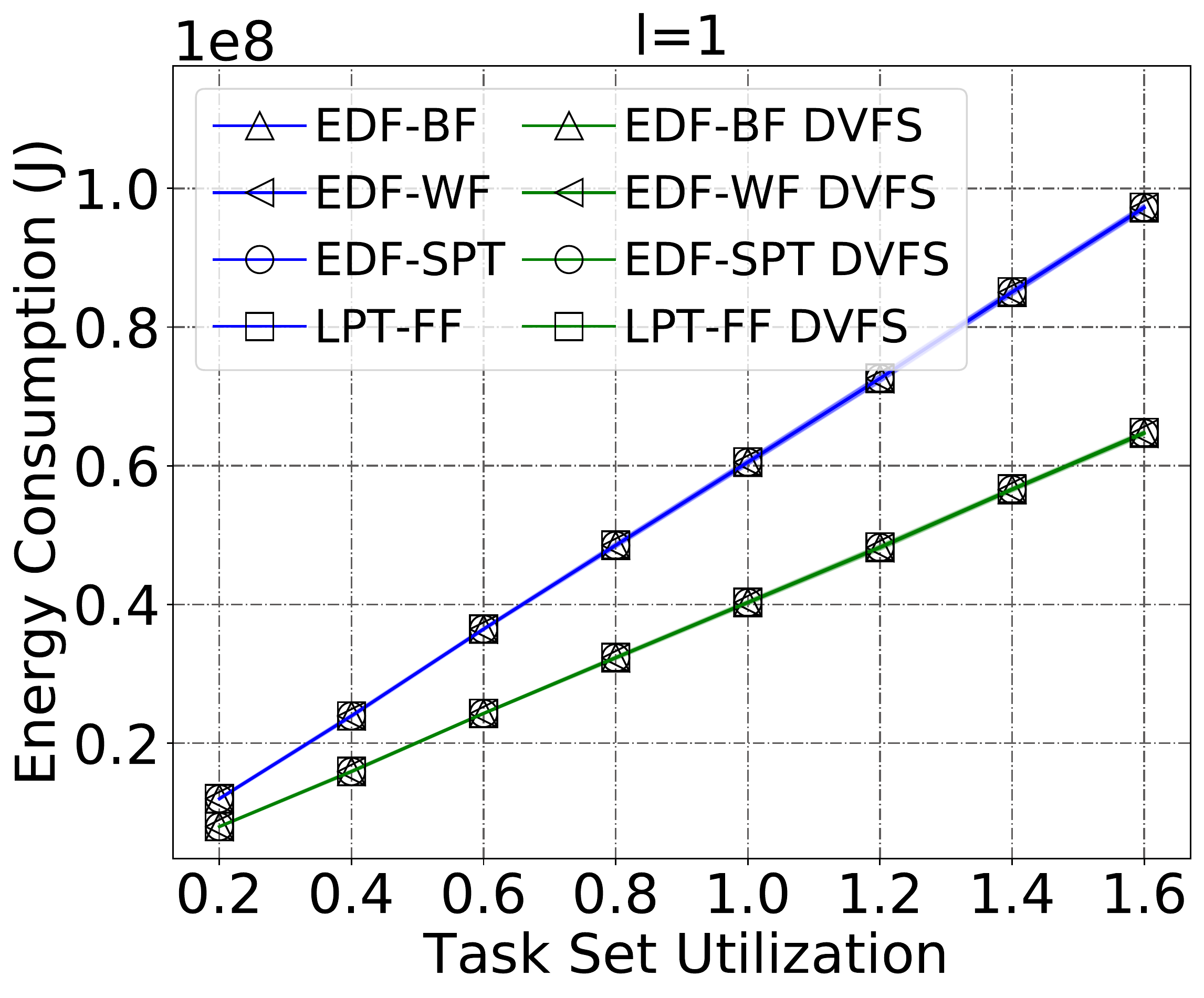}
		\label{fig:off_dvfs_E}
	}
	\subfloat[]
	{
		\includegraphics[width=0.47\linewidth]{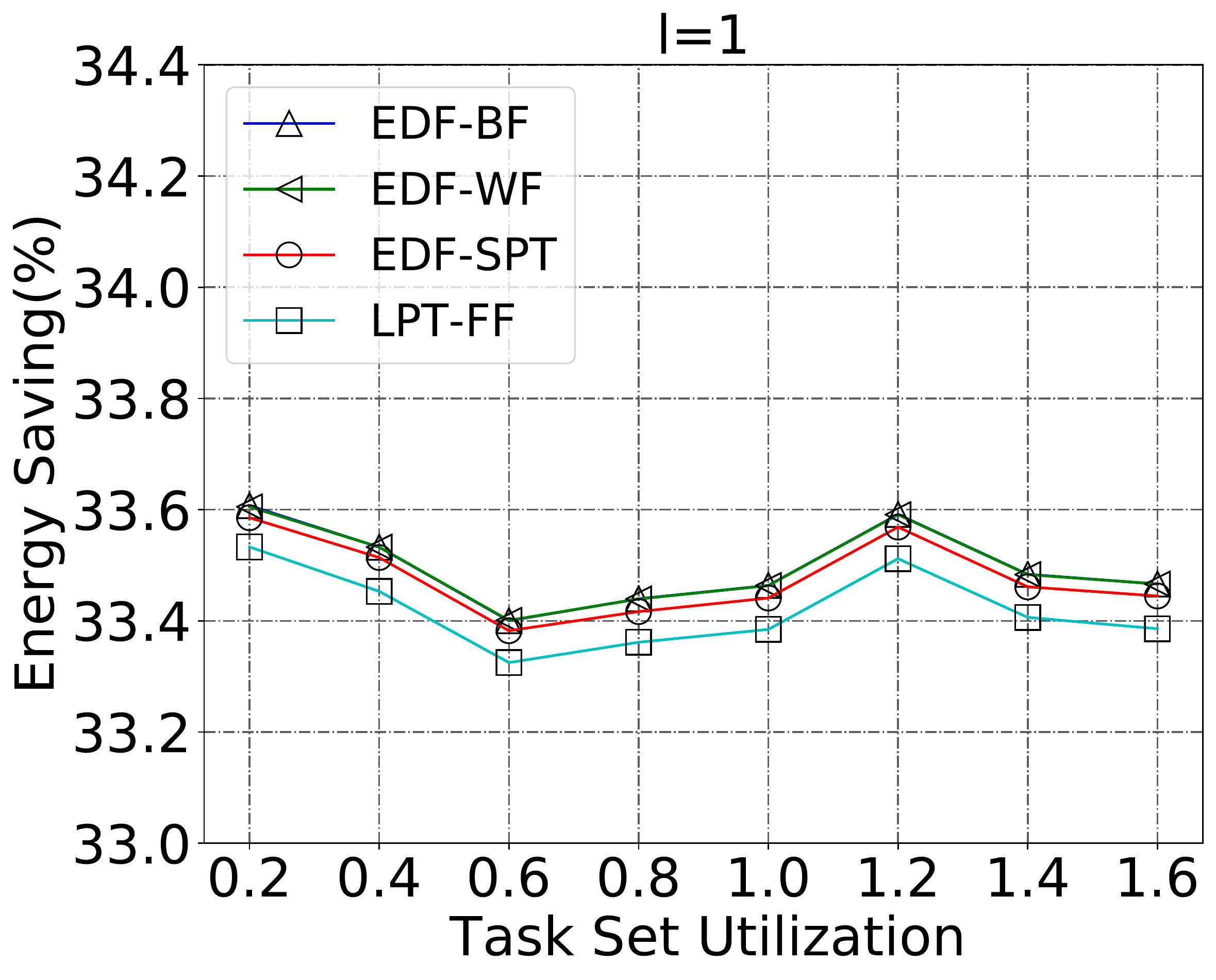}
		\label{fig:off_dvfs_saving}
	}
	\caption{The energy consumption of the non-DVFS scheduling algorithms and the DVFS scheduling algorithms when $l=1$.}\label{fig:off_dvfs}
\end{figure}

Figure \ref{fig:off_dvfs_E} shows the baseline energy consumption of different task set utilization. The four solid blue lines in Figure \ref{fig:off_dvfs_E} are overlapped, which also proves that the baseline energy is independent of the scheduling algorithm. The baseline energy increases linearly to the task set utilization. 

\begin{figure}[ht]
	\centering
	\includegraphics[width=0.47\textwidth]{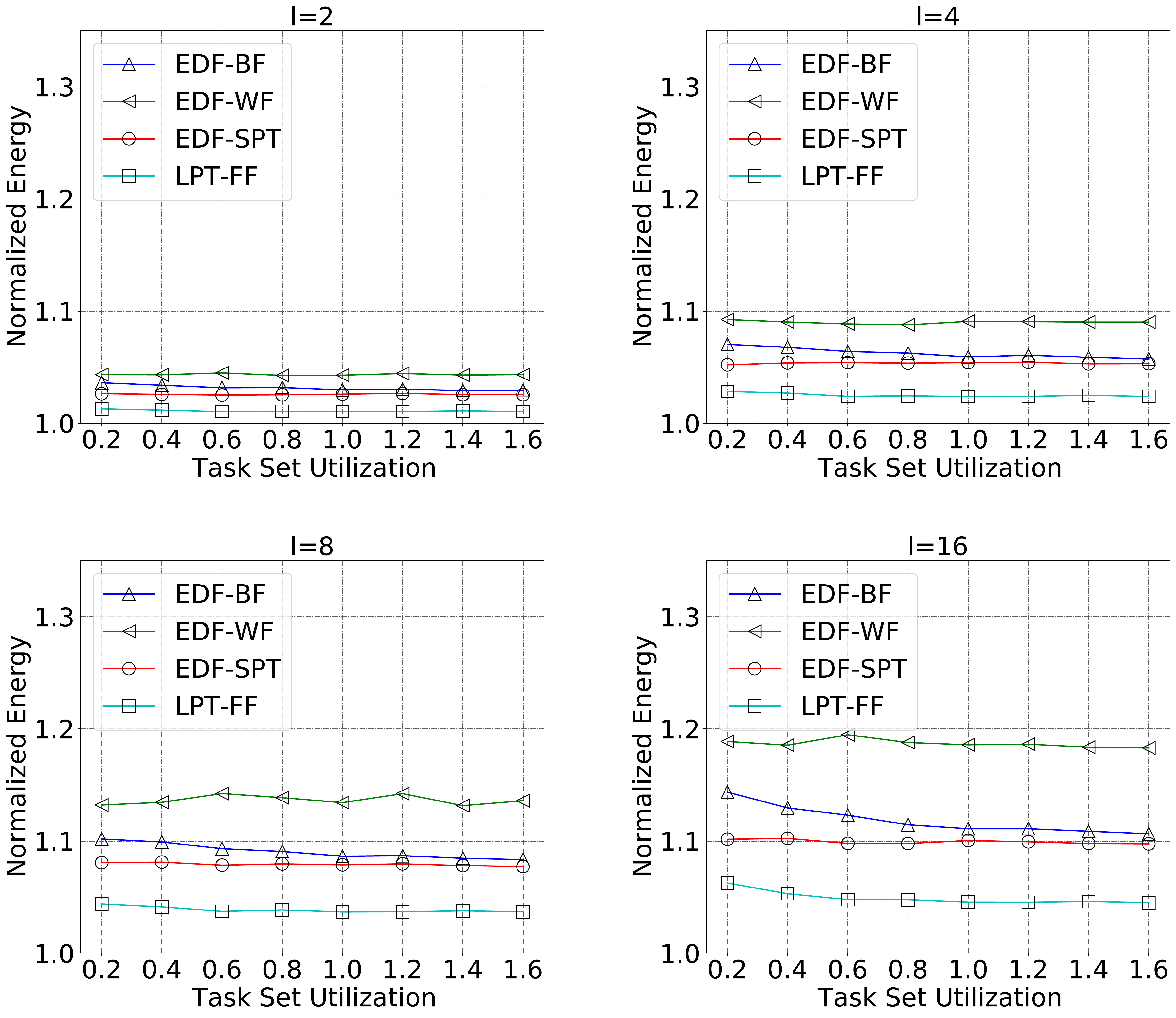}
	\caption{The energy consumption of the non-DVFS scheduling algorithms.}\label{fig:off_l}
\end{figure}

For ease of comparison, we compute the normalized energy consumption of other $l$ values (normalized to the baseline energy). The difference between the normalized energy and the baseline is caused by the idle energy consumption. Algorithms with less idle energy consumption and less servers show better performance. Figure \ref{fig:off_l} shows the normalized energy of $l>1$. 
The idle energy is non-trivial when the $U_{\textbf{J}}$ is small and $l$ is large. The LPT-FF scheduling algorithm has the highest idle system energy. When $l$ = 16 and $U_{\textbf{J}}=0.2$, LPT-FF consumes about 31\% idle energy. When $U_{\textbf{J}}$ is large, the energy consumptions converge to their baseline, where LPT-FF and EDF-WF show slow convergence speed. Our EDL algorithm has the least idle system energy, and the idle energy converges to zero fast.

\begin{figure}[ht]
	\centering
	\includegraphics[width=0.47\textwidth]{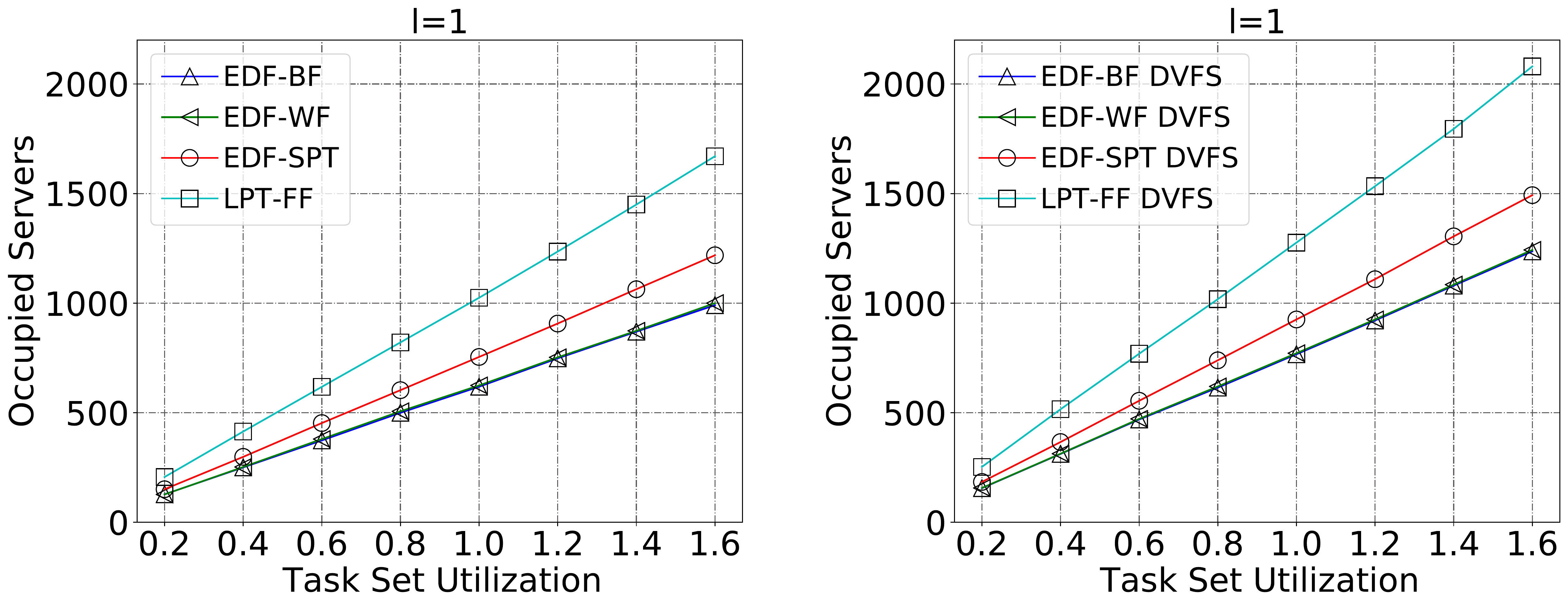}
	\caption{Comparison between non-DVFS and DVFS scheduling algorithms in the number of occupied servers when $l=1$.}\label{fig:off_nodes}
\end{figure}

Figure \ref{fig:off_nodes} shows the number of occupied servers of the schedule algorithms when $l=1$. We sort them in occupied servers descending order, as: LPT-FF, EDL, EDF-WF and EDF-BF. The number of occupied servers is also linearly to the task set utilization. For $l>1$, the number of occupied servers has similar trends, that our EDF scheduling algorithm consumes much fewer servers than LPT.

To summarize, our EDL algorithm with DVFS shows decent energy and computation resource conservation with different server modes and various task set utilization, which indicates that it is efficient in controlling the system idle energy. The LPT-FF algorithm is poor in computation resource conservation but has decent energy conservation.

\subsubsection{DVFS Performance}
We plot the number of occupied servers with $l=1$ in Figure \ref{fig:off_nodes}. It turns out that LPT-FF DVFS algorithm occupies the most servers. Our EDF DVFS algorithm still consumes much fewer server resources than LPT-FF DVFS. The performance of EDF-BF DVFS and EDF-WF DVFS are identical. 
%We also compute the normalized server numbers by dividing them by those of non-DVFS algorithms. We do not demo the results of other $l$ values since they are almost the same with that of $l=1$.

We first conduct the EDL DVFS algorithm with $\theta=1$. Analytically, the energy saving of the DVFS-based scheduling algorithm has an upper bound, which equals the average runtime energy saving of the benchmark applications. Our benchmark-based simulation task set is a combination of both energy-prior and deadline-prior tasks, and the deadline-prior tasks need to sacrifice the energy saving for the deadline so that the overall saving cannot exceed the theoretical average energy savings of the 20 benchmark applications, which is 36.4\% in our case.

We plot the absolute DVFS-based energy consumption with the broken lines in Figure \ref{fig:off_dvfs_E} and the energy saving in Figure \ref{fig:off_dvfs_saving} when $l=1$. The energy saving slightly varies around 33\%, with a mean value of 33.5\%. The about 3\% of energy saving loss compared to the theoretical upper bound is caused by the deadline-prior tasks in the task set.

\begin{figure}[ht]
	\centering
	\includegraphics[width=0.47\textwidth]{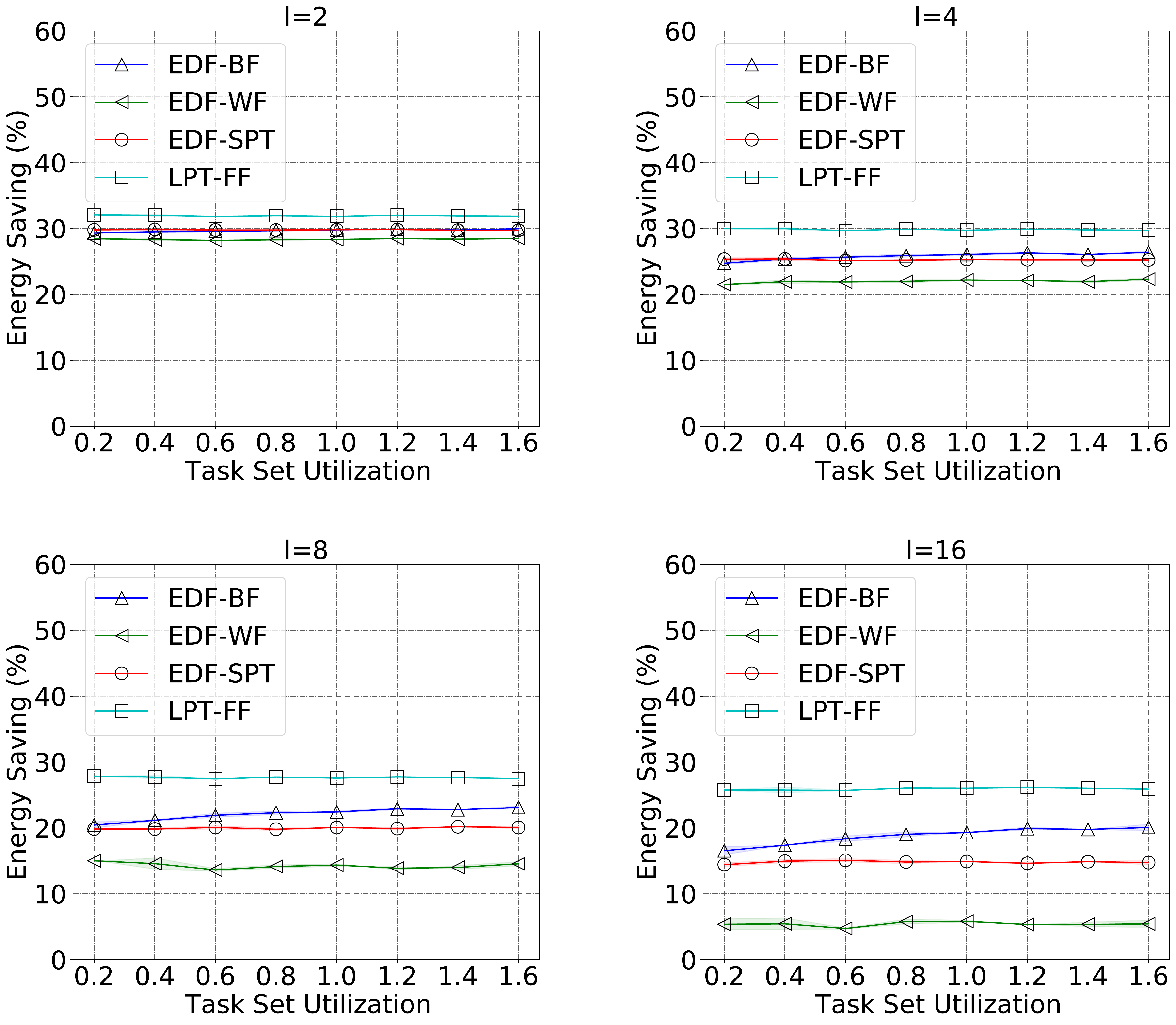}
	\caption{The energy savings of the DVFS-based algorithms compared to the baseline energy consumption when $l>1$.}\label{fig:off_saving}
\end{figure}

Figure \ref{fig:off_saving} shows the energy savings of different server modes when $l>1$. Overall speaking, small $l$ have higher energy savings/decrease due to lower idle energy consumption. The LPT-FF DVFS algorithm saves the most energy, while the EDF-WF algorithm consumes the most energy. Our EDL DVFS algorithm is quite close with the EDF-BF DVFS, but when $l$ and $U_{\textbf{J}}$ are large, it saves slightly less energy, about 5.1\% of energy decrease with $l=16$ and $U_{\textbf{J}}=1.6$, than the EDF-BF DVFS scheduling algorithm.

\subsubsection{Effectiveness of the EDL Readjustment}
\begin{figure}[ht]
	\centering
	\includegraphics[width=0.47\textwidth]{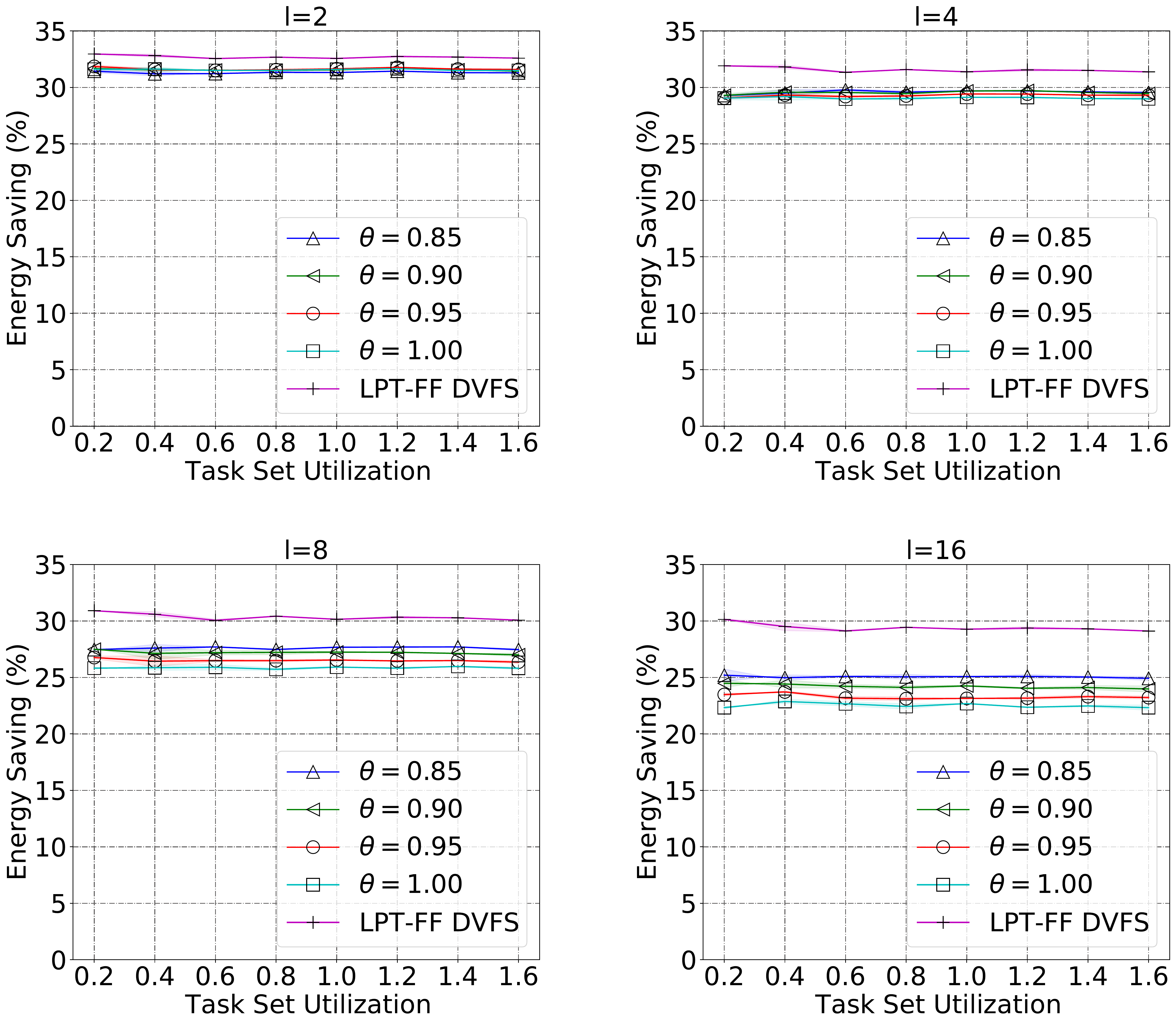}
	\caption{Energy savings of EDL $\theta$-readjustment DVFS algorithm.}\label{fig:off_theta}
\end{figure}

We then validate the EDL DVFS algorithm with other $\theta$ values. The $\theta$ is designed for reducing the system idle energy, so it is effective when $l>1$ only. We measure the system energy consumption with $\theta=0.8,0.85,0.9,0.95$ and $1$, each for 1000 times. 
%We find the minimum energy consumption and the corresponding best $\theta$ for each group of $l$ and $U_{\textbf{J}}$. 
We compute the average energy savings of each $\theta$ and compare them to those of LPT-FF DVFS, which conserves the most energy in our previous experimental results. Figure \ref{fig:off_theta} shows the results. When $l\le4$, the average energy savings of different $\theta$ are almost the same, which are no more than 3\% smaller than those of the LPT-FF DVFS algorithm. However, when $l$ increases, our $\theta$-readjustment scheme can decrease the overall idle energy and make the EDL DVFS lines approach the LPT-FF DVFS line. 
\subsection{Online EDL DVFS Performance} \label{subsec:ex_edl_on}

We compare our EDL algorithm for the online case to a classical bin-packing heuristic algorithm described in Algorithm \ref{algo:bin_packing}. The idea of bin-packing heuristic has been used in \cite{Liu2012}, and we modify their algorithm to fit our system model. The complexity of this algorithm is $n(\text{log}n + \Phi + nm)$, larger than that of the EDL algorithm, mainly caused by the frequent updates of the processor load. 

\subsubsection{Baseline Performance}
As proven in \cite{hong1992}, there is no optimal solution for our online task scheduling problem. In this work, we refer to the performance of the task scheduling algorithms without GPU DVFS as the baseline performance. In particular, we execute the EDL algorithm without runtime readjustment, i.e., $\theta = 1$.

\begin{figure}[ht]
	\centering
	\includegraphics[width=0.47\textwidth]{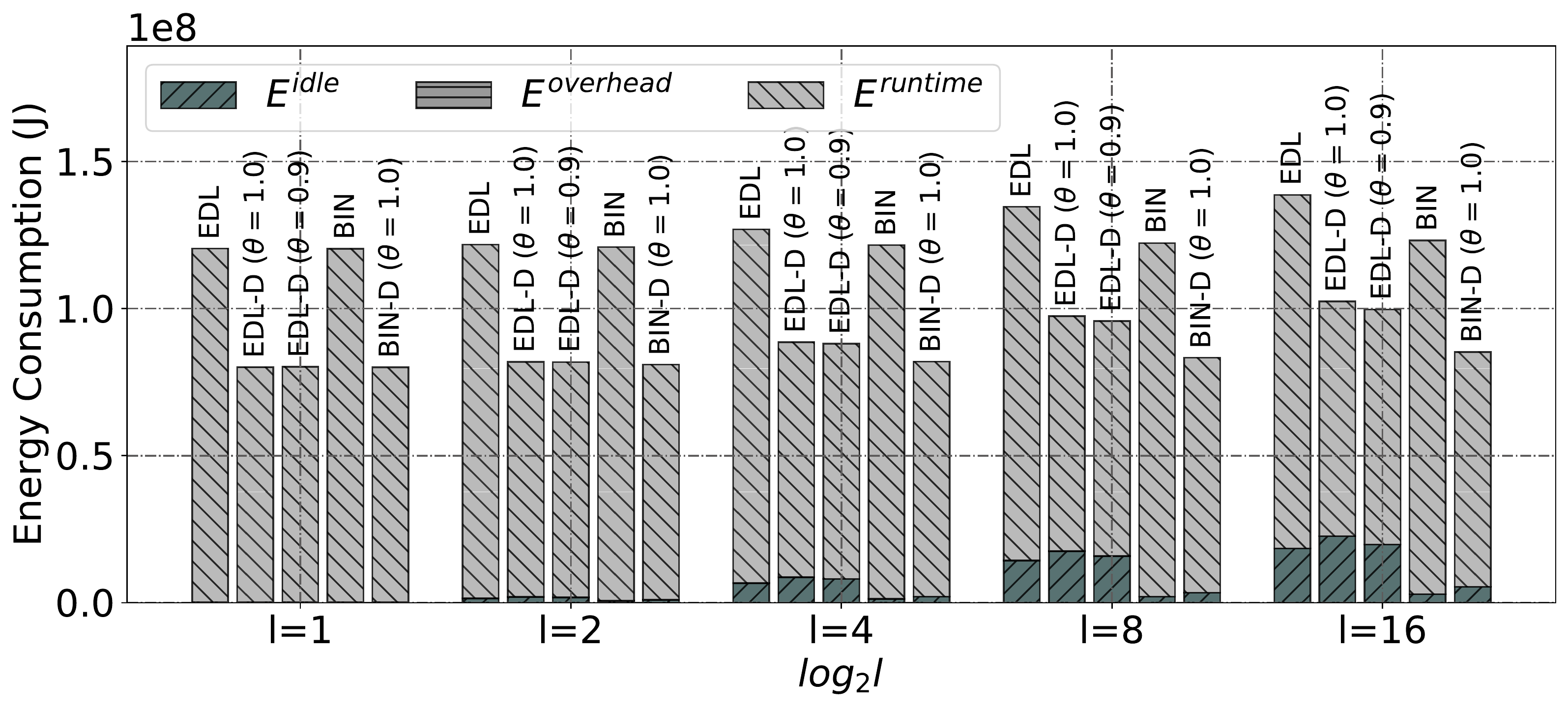}
	\caption{Decomposition of the total energy consumption. "EDL" and "BIN" denote our EDL readjustment algorithm and the bin-packing algorithm without GPU DVFS, while "EDL-D" and "BIN-D" denote the algorithms with GPU DVFS.}\label{fig:e_dvfs_bar}
\end{figure}

\begin{figure*}[htbp]
	\centering
	\includegraphics[width=0.87\textwidth]{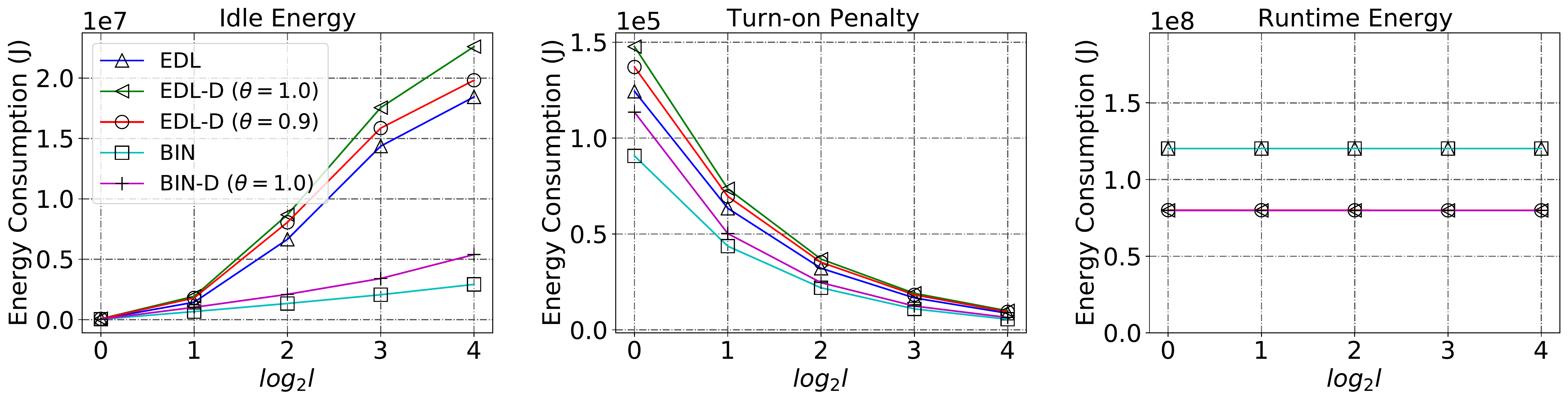}
	\caption{Comparison between the energy consumption of the non-DVFS and DVFS scheduling algorithms.}\label{fig:e_dvfs_line}
\end{figure*}

We show the total energy decomposition in Figure \ref{fig:e_dvfs_bar}, where the two highest bars denote the baseline energy consumption. The EDL algorithm has similar energy consumption to the bin-packing algorithm when $l$ is small, but leads to larger idle energy consumption when $l$ is large. The runtime energy consumption is independent of $l$ or the scheduling algorithm, with a constant value of 120.22 MJ. The overhead energy is marginal in the whole energy portfolio, varying from 5.39 KJ to 120.62 KJ, and it slightly decreases as $l$ increases. The idle energy consumption changes to the server configuration significantly, varying from 33.64 KJ to 18.44 MJ.

The larger idle energy consumption is mainly caused by those idle CPU-GPU pairs that cannot be turned off even if no active task has been assigned to them. When $l = 1$, each CPU-GPU pair is idle for at most $\rho$ after task processing, while when $l = 16$, the idle period of a CPU-GPU pair is overall much longer, that the CPU-GPU pairs on one server are idle for at least about $\rho$. Intuitively the more load balanced at runtime, the less idle system energy is. When we examine the runtime task mapping status, the bin-packing first-fit algorithm usually occupies fewer CPU-GPU pairs than EDL, which makes it consume less idle energy. 

\subsubsection{DVFS Performance}
We conduct DVFS experiments with $\theta = 1$ and $\theta = 0.9$ firstly, and then discuss the readjustment with other values of $\theta$. For each group of experiments, we use the same offline and online task sets as those of the baseline simulation.

Figure \ref{fig:e_dvfs_bar} shows the DVFS energy consumption with three lower bars. The runtime energy consumption of the DVFS algorithms is still a constant. It reduces from 122.22 MJ to 79.87 MJ; about 34.7\% of runtime energy is saved with GPU DVFS. When $l$ = 1, the three algorithms have similar energy consumption, about 80.03-80.24 MJ, where the bin-packing DVFS algorithm is slightly better. For other values of $l$, the EDL readjustment ($\theta = 0.9$) DVFS algorithm consumes less total energy than the EDL DVFS ($\theta = 1$) algorithm. When $l$ = 16, the total energy consumption of the three algorithms are 102.49, 99.73, 85.27 MJ respectively.

We further compare the idle energy and the turn-on overhead in Figure \ref{fig:e_dvfs_line}. The DVFS algorithms lead to increases of idle system energy, especially for the EDL DVFS algorithm without runtime adjustment. If the runtime $\theta$-readjustment is applied, the idle energy is effectively controlled. When $l$ = 16, the idle energy of the EDL non-DVFS, DVFS without readjustment and DVFS $\theta$-readjustment algorithms are 18.44, 22.61 and 19.82 MJ, respectively. 
%For the bin-packing algorithm, the difference between the DVFS and non-DVFS in idle energy consumption is relatively small, as 0.22 GJ when $l$ = 16. 
The turn-on overhead is still marginal in the whole energy portfolio. In general, the bin-packing algorithm is more effective in controlling the turn-on overhead, which means that the newly arriving tasks are more likely to be assigned to current busy servers, while it is the opposite for the EDL DVFS algorithm without readjustment.

To summarize, the bin-packing first-fit algorithm with the EDF order has better performance in the energy conservation in both the baseline and the DVFS simulation. Besides, a runtime readjustment is needed when a server has many CPU-GPU pairs. A better balance between the runtime energy and the idle energy \& turn-on overhead is preferred when GPU DVFS is applied.

\subsubsection{Effectiveness of the EDL Readjustment}
In the previous section, we have confirmed that the $\theta$-readjustment is effective in controlling the idle energy when GPU DVFS is applied. We now discuss the impact of $\theta$ on the effectiveness of the readjustment strategy.

\begin{figure}[ht]
	\centering
	\includegraphics[width=0.46\textwidth]{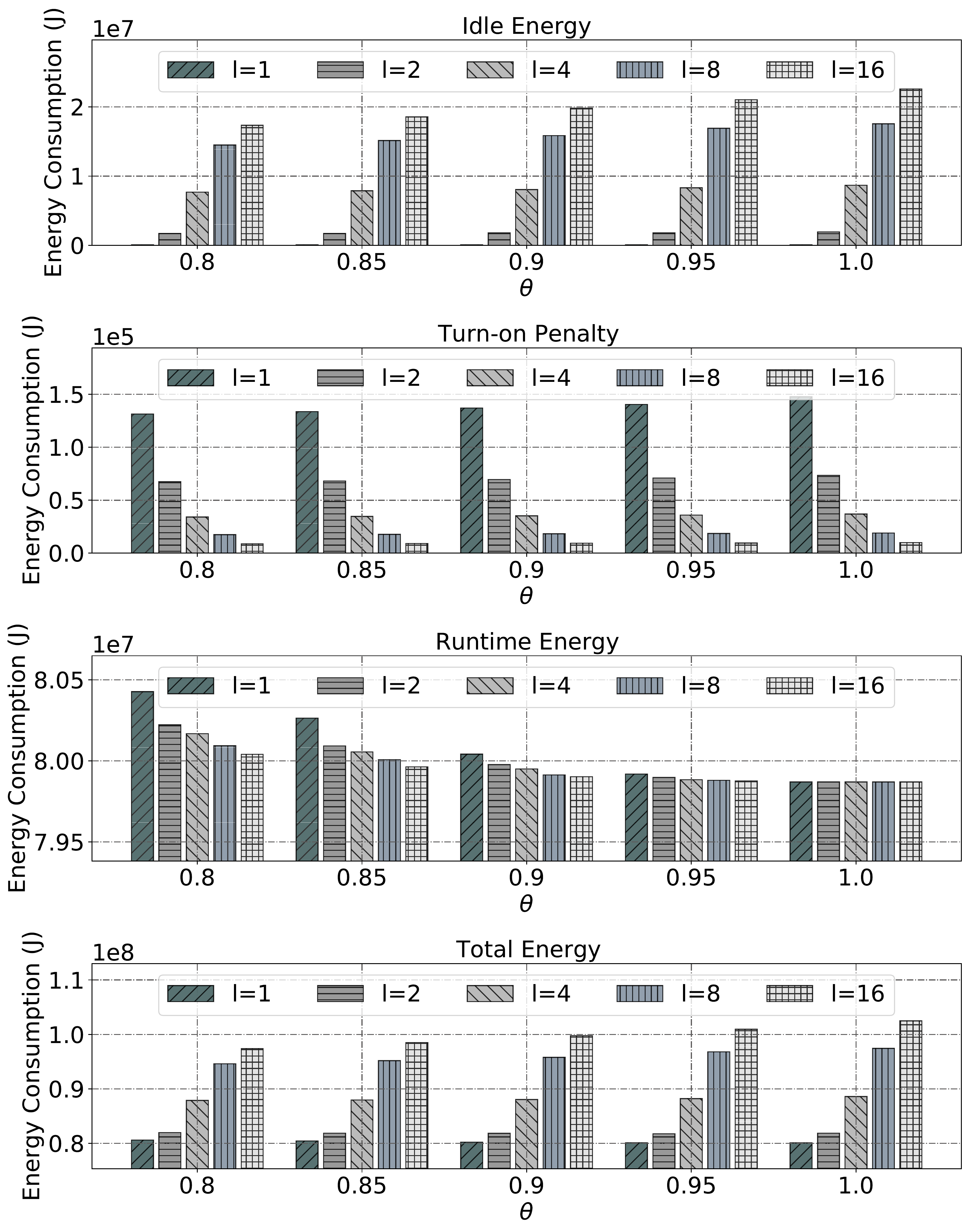}
	\caption{The energy consumption with runtime readjustments.}\label{fig:e_dvfs_theta}
\end{figure}
We conduct the EDL DVFS $\theta$-readjustment algorithm with five different values of $\theta$. We plot the average idle energy, turn-on overhead, runtime energy and the total energy in Figure \ref{fig:e_dvfs_theta}. It is clear from our experimental results that smaller $\theta$ will result in slightly larger runtime energy consumption but less idle energy and turn-on energy. With $\theta \neq 1$, we consume less total energy, less idle energy and less turn-on overhead, especially for large $l$. For example, when $l = 16$, applying $\theta = 0.95$ reduces the total energy consumption from 102.49 MJ to 100.94 MJ, and reduces the idle energy from 22.61 MJ to 21.05 MJ. It is notable that when $\theta \neq 1$, the turn-on overhead do not vary much to $\theta$ for the same $l$, while the idle energy decreases with a smaller $\theta$ and a larger $l$, and the runtime energy increases with a smaller $\theta$ and a smaller $l$. Much more energy is consumed when $\theta = 1$, therefore a runtime readjustment is quite necessary. For all the experiments, $\theta = 0.8$ ends in the minimum total energy consumption only except $l = 1$.

\begin{figure}[ht]
	\centering
	\includegraphics[width=0.47\textwidth]{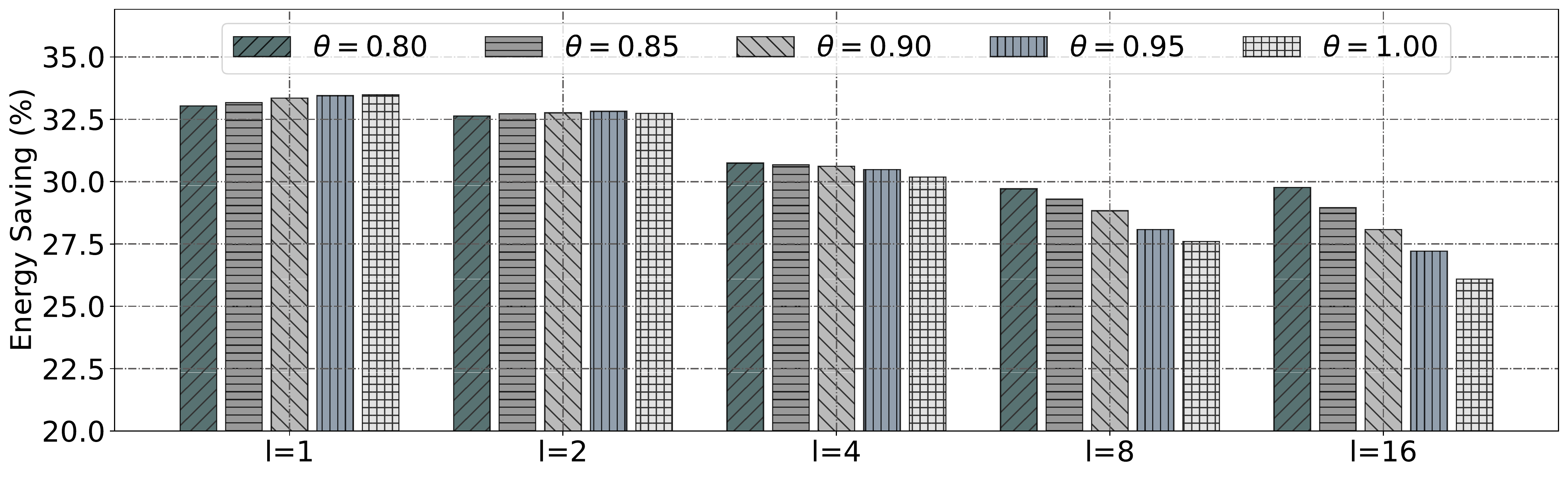}
	\caption{The energy reduction compared to the baseline energy consumption.}\label{fig:online_saving}
\end{figure}

Figure \ref{fig:online_saving} shows the energy reduction compared to the baseline total energy consumption of the EDL algorithm of all the $\theta$ configurations. Theoretically speaking, the energy reduction has an upper bound which is the average runtime energy reduction of the set of benchmark applications, i.e. 35\% in our case. With an appropriate $\theta$, our online EDL algorithm can conserve 30-33\% of energy, almost comparable to the theoretical upper limit. But as $l$ becomes larger, the energy reduction gradually decreases due to the considerable idle energy. Besides, the energy conservation of larger $l$ depends on the $\theta$-readjustment more strongly. The selection of parameter $\theta$ depends on the ratio of the runtime energy over the idle energy. Our simulation results show that setting $\theta$ to 0.8 generally guarantees better energy conservation for those servers that are equipped with more CPU-GPU pairs.
%Fig. 7 shows the energy reduction compared to the baseline total energy consumption of the EDL algorithm of all the $\theta$ configurations. Theoretically the energy reduction is upper bounded by the average runtime energy reduction of the set of benchmark applications, i.e., 38\% in our case. With appropriate $\theta$, our online EDL algorithm can conserve 30-36\% of energy, very close to the theoretical upper limit. But as $l$ becomes larger, the energy reduction gradually decreases. Besides, the energy conservation of larger $l$ depends more on the $\theta$-readjustment. The selection of parameter $\theta$ depends on the ratio of the runtime energy over the idle energy.

\section{Conclusion} \label{sec:conclusion}

In this paper, we study the energy conserving problem on CPU-GPU hybrid clusters. We propose the GPU-specific DVFS power and performance models, and derive the appropriate GPU voltage/frequency setting through the mathematical optimization. We also design a heuristic scheduling algorithm for both the offline and online modes to assign multiple tasks to the cluster, which uses the runtime DVFS readjustment to make a good balance between the dynamic energy consumption and static energy consumption. We find that for both the offline and online tasks, the static energy is non-negligible, that a better balance of the dynamic and static energy is quite necessary. 
Our $\theta$-readjustment EDL algorithm can reserve the advantage of GPU DVFS and avoid large idle power overhead caused by low utilization of those servers with multiple CPU-GPU pairs.

In this work, we make a number of assumptions in the problem formulation to simplify the problem, such as homogeneity of CPUs and GPUs. We leave a more practical formulation and solution for our future work. It is also interesting to consider the case that a single task can occupy multiple GPUs, which is a typical case of modern distributed deep learning applications.
	
\section*{Acknowledgements}
This work was supported in part by the Hong Kong RGC GRF grant under the contract HKBU 12200418 and grant RMGS2019\_1\_23 71 from Hong Kong Research Matching Grant Scheme.
	
\bibliographystyle{IEEEtran}
\bibliography{main.bbl}

\begin{thebibliography}{10}
\providecommand{\url}[1]{#1}
\csname url@samestyle\endcsname
\providecommand{\newblock}{\relax}
\providecommand{\bibinfo}[2]{#2}
\providecommand{\BIBentrySTDinterwordspacing}{\spaceskip=0pt\relax}
\providecommand{\BIBentryALTinterwordstretchfactor}{4}
\providecommand{\BIBentryALTinterwordspacing}{\spaceskip=\fontdimen2\font plus
\BIBentryALTinterwordstretchfactor\fontdimen3\font minus
  \fontdimen4\font\relax}
\providecommand{\BIBforeignlanguage}[2]{{%
\expandafter\ifx\csname l@#1\endcsname\relax
\typeout{** WARNING: IEEEtran.bst: No hyphenation pattern has been}%
\typeout{** loaded for the language `#1'. Using the pattern for}%
\typeout{** the default language instead.}%
\else
\language=\csname l@#1\endcsname
\fi
#2}}
\providecommand{\BIBdecl}{\relax}
\BIBdecl

\bibitem{googleAI2016}
J.~Clark, ``{Google cuts its giant electricity bill with DeepMind-powered
  AI},'' [Online]
  https://www.datacenterknowledge.com/\\archives/2016/07/19/google-cuts-its-giant-electricity-bill-with-deepmind-powered-ai,
  Google.

\bibitem{silver2016mastering}
D.~Silver, A.~Huang, C.~J. Maddison, A.~Guez, L.~Sifre, G.~Van Den~Driessche,
  J.~Schrittwieser, I.~Antonoglou, V.~Panneershelvam, M.~Lanctot \emph{et~al.},
  ``Mastering the game of go with deep neural networks and tree search,''
  \emph{nature}, vol. 529, no. 7587, p. 484, 2016.

\bibitem{floridi2020gpt}
L.~Floridi and M.~Chiriatti, ``Gpt-3: Its nature, scope, limits, and
  consequences,'' \emph{Minds and Machines}, vol.~30, no.~4, pp. 681--694,
  2020.

\bibitem{xin2021automl}
X.~He, K.~Zhao, and X.~Chu, ``Automl: A survey of the state-of-the-art,''
  \emph{Knowledge-Based Systems}, vol. 212, p. 106622, 2021.

\bibitem{powerS2016}
R.~A. Bridges, N.~Imam, and T.~M. Mintz, ``Understanding gpu power: A survey of
  profiling, modeling, and simulation methods,'' \emph{ACM Computing Surveys},
  vol.~49, no.~3, p.~27, Sep. 2016.

\bibitem{survey2017}
X.~Mei, Q.~Wang, and X.~Chu, ``A survey and measurement study of gpu dvfs on
  energy conservation,'' \emph{Digital Communications and Networks}, vol.~3,
  no.~2, pp. 89 -- 100, 2017.

\bibitem{top500}
E.~Strohmaier, J.~Dongarra, H.~Simon, and M.~Meuer, ``{TOP500},'' [Online]
  http://www.top500.org, TOP500.

\bibitem{green500}
W.~FENG and T.~SCOGLAND, ``{The Green500 list, November, 2020},'' [Online]
  https://www.top500.org/green500/lists/2020/11/, Green500.

\bibitem{mei2014benchmark}
X.~Mei, K.~Zhao, C.~Liu, and X.~Chu, ``Benchmarking the memory hierarchy of
  modern gpus,'' in \emph{IFIP International Conference on Network and Parallel
  Computing}, 2014.

\bibitem{mei2017tpds}
X.~{Mei} and X.~{Chu}, ``Dissecting gpu memory hierarchy through
  microbenchmarking,'' \emph{IEEE Transactions on Parallel and Distributed
  Systems}, vol.~28, no.~1, pp. 72--86, Jan 2017.

\bibitem{tang2019dvfs_dl}
Z.~Tang, Y.~Wang, Q.~Wang, and X.~Chu, ``The impact of gpu dvfs on the energy
  and performance of deep learning: An empirical study,'' in \emph{Proceedings
  of the Tenth ACM International Conference on Future Energy Systems}, ser.
  e-Energy '19.\hskip 1em plus 0.5em minus 0.4em\relax New York, NY, USA:
  Association for Computing Machinery, 2019, pp. 315--325.

\bibitem{fan2007}
X.~Fan, W.-D. Weber, and L.~A. Barroso, ``Power provisioning for a
  warehouse-sized computer,'' in \emph{Proceedings of the 34th Annual
  International Symposium on Computer Architecture}, ser. ISCA '07, 2007, pp.
  13--23.

\bibitem{isci2006}
C.~Isci, A.~Buyuktosunoglu, A.~Buyuktosunoglu, C.-Y. Cher, P.~Bose, and
  M.~Martonosi, ``An analysis of efficient multi-core global power management
  policies: Maximizing performance for a given power budget,'' in
  \emph{Proceedings of the 39th Annual IEEE/ACM International Symposium on
  Microarchitecture}, ser. MICRO 39, 2006, pp. 347--358.

\bibitem{ge2013}
R.~{Ge}, R.~{Vogt}, J.~{Majumder}, A.~{Alam}, M.~{Burtscher}, and Z.~{Zong},
  ``Effects of dynamic voltage and frequency scaling on a k20 gpu,'' in
  \emph{Proceedings of the 42nd International Conference on Parallel
  Processing}, Oct 2013, pp. 826--833.

\bibitem{yuki2012}
Y.~Abe, H.~Sasaki, M.~Peres, K.~Inoue, K.~Murakami, and S.~Kato, ``Power and
  performance analysis of gpu-accelerated systems,'' in \emph{Proceedings of
  the 2012 Workshop on Power-Aware Computing and Systems}.\hskip 1em plus 0.5em
  minus 0.4em\relax Hollywood, CA: {USENIX}, 2012.

\bibitem{yuki2014}
Y.~{Abe}, H.~{Sasaki}, S.~{Kato}, K.~{Inoue}, M.~{Edahiro}, and M.~{Peres},
  ``Power and performance characterization and modeling of gpu-accelerated
  systems,'' in \emph{Proceedings of the 28th IEEE International Parallel and
  Distributed Processing Symposium}, May 2014, pp. 113--122.

\bibitem{wang2020dvfs}
Q.~{Wang} and X.~{Chu}, ``Gpgpu performance estimation with core and memory
  frequency scaling,'' \emph{IEEE Transactions on Parallel and Distributed
  Systems}, vol.~31, no.~12, pp. 2865--2881, 2020.

\bibitem{racing2015}
D.~H.~K. {Kim}, C.~{Imes}, and H.~{Hoffmann}, ``Racing and pacing to idle:
  Theoretical and empirical analysis of energy optimization heuristics,'' in
  \emph{2015 IEEE 3rd International Conference on Cyber-Physical Systems,
  Networks, and Applications}, Aug 2015, pp. 78--85.

\bibitem{xin2017schedule}
X.~{Mei}, X.~{Chu}, H.~{Liu}, Y.~{Leung}, and Z.~{Li}, ``Energy efficient
  real-time task scheduling on cpu-gpu hybrid clusters,'' in \emph{Proceedings
  of IEEE Conference on Computer Communications}, 2017, pp. 1--9.

\bibitem{anton2012}
A.~Beloglazov, J.~Abawajy, and R.~Buyya, ``Energy-aware resource allocation
  heuristics for efficient management of data centers for cloud computing,''
  \emph{Future Generation Computer Systems}, vol.~28, no.~5, pp. 755 -- 768,
  2012, special Section: Energy efficiency in large-scale distributed systems.

\bibitem{tang2016energy}
Z.~Tang, L.~Qi, Z.~Cheng, K.~Li, S.~U. Khan, and K.~Li, ``An energy-efficient
  task scheduling algorithm in dvfs-enabled cloud environment,'' \emph{Journal
  of Grid Computing}, vol.~14, no.~1, pp. 55--74, 2016.

\bibitem{fan2019predictable}
K.~Fan, B.~Cosenza, and B.~Juurlink, ``{Predictable GPUs Frequency Scaling for
  Energy and Performance},'' in \emph{Proceedings of the 48th ICPP,
  2019}.\hskip 1em plus 0.5em minus 0.4em\relax IEEE, 2019.

\bibitem{dvfs2019}
J.~Guerreiro, A.~Ilic, N.~Roma, and P.~Tomás, ``Dvfs-aware application
  classification to improve gpgpus energy efficiency,'' \emph{Parallel
  Computing}, vol.~83, pp. 93 -- 117, 2019.

\bibitem{huang2019}
Y.~{Huang}, B.~{Guo}, and Y.~{Shen}, ``Gpu energy consumption optimization with
  a global-based neural network method,'' \emph{IEEE Access}, vol.~7, pp.
  64\,303--64\,314, 2019.

\bibitem{ali2015}
A.~Karami, F.~Khunjush, and S.~A. Mirsoleimani, ``{A statistical performance
  analyzer framework for OpenCL kernels on Nvidia GPUs},'' \emph{J.
  Supercomput.}, vol.~71, no.~8, pp. 2900--2921, Aug. 2015.

\bibitem{wu2015gpgpu}
G.~Wu \emph{et~al.}, ``{GPGPU performance and power estimation using machine
  learning},'' in \emph{Proceedings of the 21st IEEE International Symposium on
  High Performance Computer Architecture}, 2015, pp. 564--576.

\bibitem{wong2010demystifying}
H.~{Wong}, M.~{Papadopoulou}, M.~{Sadooghi-Alvandi}, and A.~{Moshovos},
  ``Demystifying gpu microarchitecture through microbenchmarking,'' in
  \emph{Proceedings of IEEE International Symposium on Performance Analysis of
  Systems Software (ISPASS)}, 2010, pp. 235--246.

\bibitem{hong2009analytical}
S.~Hong and H.~Kim, ``{An analytical model for a GPU architecture with
  memory-level and thread-level parallelism awareness},'' in \emph{Proceedings
  of the 36th ISCA}, ser. ISCA '09.\hskip 1em plus 0.5em minus 0.4em\relax ACM,
  2009.

\bibitem{song2013simplified}
S.~Song, C.~Su, B.~Rountree, and K.~W. Cameron, ``{A simplified and accurate
  model of power-performance efficiency on emergent GPU architectures},'' in
  \emph{Proceedings of the 27th IPDPS, 2013}.

\bibitem{nath2015crisp}
R.~Nath and D.~Tullsen, ``{The CRISP performance model for dynamic voltage and
  frequency scaling in a GPGPU},'' in \emph{Proceedings of the 48th MICRO,
  2015}.

\bibitem{Vignesh2016}
V.~Adhinarayanan, B.~Subramaniam, and W.~Feng, ``Online power estimation of
  graphics processing units,'' in \emph{Proceedings of the 16th IEEE/ACM
  International Symposium on Cluster, Cloud and Grid Computing (CCGrid)}, May
  2016, pp. 245--254.

\bibitem{Dutta2018}
B.~Dutta, V.~Adhinarayanan, and W.-c. Feng, ``{GPU power prediction via
  ensemble machine learning for DVFS space exploration},'' in \emph{Proceedings
  of the 15th ACM International Conference on Computing Frontiers (CF)}.\hskip
  1em plus 0.5em minus 0.4em\relax New York, NY, USA: ACM, 2018, pp. 240--243.

\bibitem{gpupower2018}
J.~{Guerreiro}, A.~{Ilic}, N.~{Roma}, and P.~{Tomas}, ``Gpgpu power modeling
  for multi-domain voltage-frequency scaling,'' in \emph{Proceedings of IEEE
  International Symposium on High Performance Computer Architecture (HPCA)},
  Feb 2018, pp. 789--800.

\bibitem{yao1995}
F.~{Yao}, A.~{Demers}, and S.~{Shenker}, ``A scheduling model for reduced cpu
  energy,'' in \emph{Proceedings of IEEE 36th Annual Foundations of Computer
  Science}, Oct 1995, pp. 374--382.

\bibitem{aydin2003}
H.~{Aydin} and {Qi Yang}, ``Energy-aware partitioning for multiprocessor
  real-time systems,'' in \emph{Proceedings International Parallel and
  Distributed Processing Symposium}, April 2003.

\bibitem{Albers2014}
S.~Albers, F.~M{\"u}ller, and S.~Schmelzer, ``Speed scaling on parallel
  processors,'' \emph{Algorithmica}, vol.~68, no.~2, pp. 404--425, Feb 2014.

\bibitem{hong1992}
K.~S. {Hong} and J.~Y.~. {Leung}, ``On-line scheduling of real-time tasks,''
  \emph{IEEE Transactions on Computers}, vol.~41, no.~10, pp. 1326--1331, Oct
  1992.

\bibitem{Irani2007}
S.~Irani, S.~Shukla, and R.~Gupta, ``Algorithms for power savings,'' \emph{ACM
  Transactions on Algorithms}, vol.~3, no.~4, Nov. 2007.

\bibitem{Gharaibeh2013}
A.~Gharaibeh, E.~Santos-Neto, L.~B.~a. Costa, and M.~Ripeanu, ``The energy case
  for graph processing on hybrid cpu and gpu systems,'' in \emph{Proceedings of
  the 3rd Workshop on Irregular Applications: Architectures and Algorithms},
  ser. IA3 '13.\hskip 1em plus 0.5em minus 0.4em\relax New York, NY, USA: ACM,
  2013, pp. 2:1--2:8.

\bibitem{liu2011}
W.~{Liu}, Z.~{Du}, Y.~{Xiao}, D.~A. {Bader}, and C.~{Xu}, ``A waterfall model
  to achieve energy efficient tasks mapping for large scale gpu clusters,'' in
  \emph{Proceedings of IEEE International Symposium on Parallel and Distributed
  Processing Workshops}, May 2011, pp. 82--92.

\bibitem{Liu2012}
C.~Liu, J.~Li, W.~Huang, J.~Rubio, E.~Speight, and X.~Lin, ``Power-efficient
  time-sensitive mapping in heterogeneous systems,'' in \emph{Proceedings of
  the 21st International Conference on Parallel Architectures and Compilation
  Techniques}, 2012, pp. 23--32.

\bibitem{xie2017schedule}
G.~{Xie}, G.~{Zeng}, X.~{Xiao}, R.~{Li}, and K.~{Li}, ``Energy-efficient
  scheduling algorithms for real-time parallel applications on heterogeneous
  distributed embedded systems,'' \emph{IEEE Transactions on Parallel and
  Distributed Systems}, vol.~28, no.~12, pp. 3426--3442, 2017.

\bibitem{deng2020tc}
Z.~{Deng}, Z.~{Yan}, H.~{Huang}, and H.~{Shen}, ``Energy-aware task scheduling
  on heterogeneous computing systems with time constraint,'' \emph{IEEE
  Access}, vol.~8, pp. 23\,936--23\,950, 2020.

\bibitem{vincent2017energy}
V.~Chau, X.~Chu, H.~Liu, and Y.-W. Leung, ``Energy efficient job scheduling
  with dvfs for cpu-gpu heterogeneous systems,'' in \emph{Proceedings of the
  Eighth International Conference on Future Energy Systems}, 2017, p. 1–11.

\bibitem{connor2018icpp}
C.~Imes, S.~Hofmeyr, and H.~Hoffmann, ``Energy-efficient application resource
  scheduling using machine learning classifiers,'' in \emph{Proceedings of the
  47th International Conference on Parallel Processing}, 2018.

\bibitem{salami2021toc}
B.~{Salami}, H.~{Noori}, and M.~{Naghibzadeh}, ``Fairness-aware energy
  efficient scheduling on heterogeneous multi-core processors,'' \emph{IEEE
  Transactions on Computers}, vol.~70, no.~1, pp. 72--82, 2021.

\bibitem{Nath2015}
R.~Nath and D.~Tullsen, ``The crisp performance model for dynamic voltage and
  frequency scaling in a gpgpu,'' in \emph{Proceedings of the 48th
  International Symposium on Microarchitecture}, ser. MICRO-48.\hskip 1em plus
  0.5em minus 0.4em\relax ACM, 2015, pp. 281--293.

\bibitem{aydin2001}
H.~{Aydin}, R.~{Melhem}, D.~{Mosse}, and P.~{Mejia-Alvarez}, ``Dynamic and
  aggressive scheduling techniques for power-aware real-time systems,'' in
  \emph{Proceedings 22nd IEEE Real-Time Systems Symposium (RTSS 2001) (Cat.
  No.01PR1420)}, Dec 2001, pp. 95--105.

\bibitem{cudasdk100}
NVIDIA, ``{CUDA SDK 10.0},'' [Online]
  https://developer.nvidia.com/cuda-10.0-download-archive.

\bibitem{che2009rodinia}
S.~Che \emph{et~al.}, ``Rodinia: A benchmark suite for heterogeneous
  computing,'' in \emph{Proceedings of IEEE International Symposium on Workload
  Characterization (IISWC)}.\hskip 1em plus 0.5em minus 0.4em\relax IEEE, 2009,
  pp. 44--54.

\bibitem{hong2010}
S.~Hong and H.~Kim, ``An integrated gpu power and performance model,'' in
  \emph{ACM SIGARCH Computer Architecture News}, vol.~38, no.~3.\hskip 1em plus
  0.5em minus 0.4em\relax ACM, 2010, pp. 280--289.

\bibitem{wang2018}
Q.~{Wang} and X.~{Chu}, ``Gpgpu performance estimation with core and memory
  frequency scaling,'' in \emph{Proceedings of the 24th IEEE International
  Conference on Parallel and Distributed Systems (ICPADS)}, Dec 2018, pp.
  417--424.

\bibitem{Ibarra1977}
O.~H. Ibarra and C.~E. Kim, ``Heuristic algorithms for scheduling independent
  tasks on nonidentical processors,'' \emph{J. ACM}, vol.~24, no.~2, pp.
  280--289, Apr. 1977.

\bibitem{Liu1973}
C.~L. Liu and J.~W. Layland, ``Scheduling algorithms for multiprogramming in a
  hard-real-time environment,'' \emph{J. ACM}, vol.~20, no.~1, pp. 46--61, Jan.
  1973.

\bibitem{lopez2004utilization}
J.~M. L{\'o}pez, J.~L. D{\'\i}az, and D.~F. Garc{\'\i}a, ``Utilization bounds
  for edf scheduling on real-time multiprocessor systems,'' \emph{Real-Time
  Systems}, vol.~28, no.~1, pp. 39--68, 2004.

\end{thebibliography}
	
\begin{IEEEbiography}
	[{\includegraphics[width=1in,height=1.25in,clip,keepaspectratio]
		{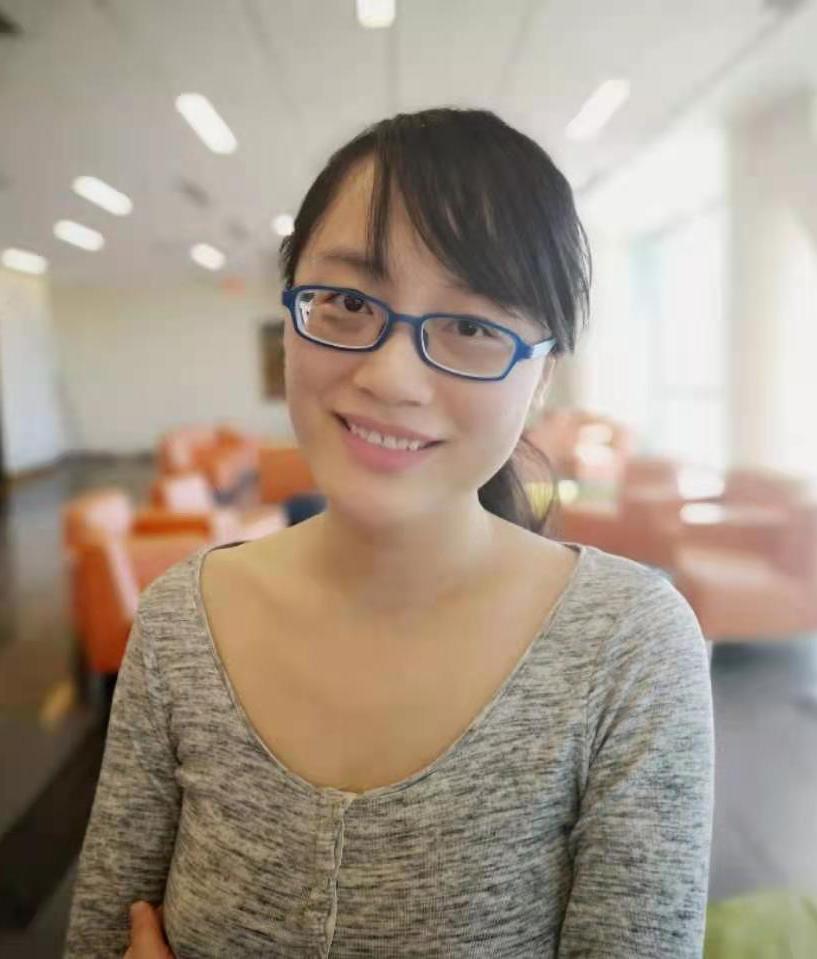}}]{Xinxin Mei}
	is a software engineer at Science Systems and Applications, Inc. Dr. Mei received her B.E. degree in electronic information engineering from the University of Science and Technology of China, P.R. China, in 2010, and the Ph.D. degree in computer science from Hong Kong Baptist University in 2016. Her research interests include distributed and parallel computing and GPU-accelerated parallel partial differential equation solvers.
\end{IEEEbiography}
\vspace{-4.0em}
\begin{IEEEbiography}
	[{\includegraphics[width=1in,height=1.25in,clip,keepaspectratio]
		{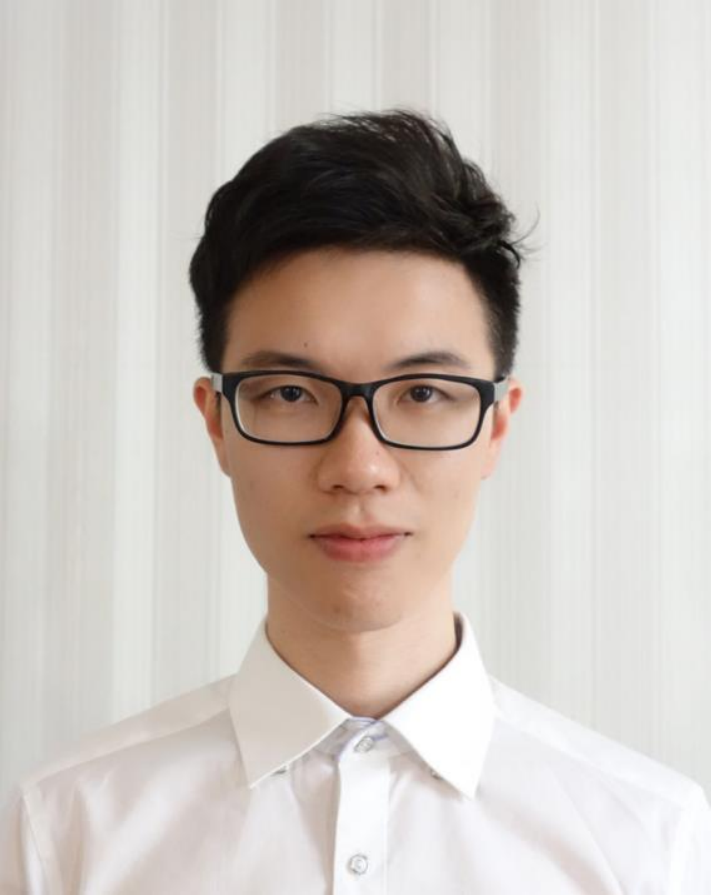}}]{Qiang Wang}
	is a research assistant professor in the Department of Computer Science, Hong Kong Baptist University. Dr. Wang received his B.E. degree from South China University of Technology in 2014, and the Ph.D. degree in computer science from Hong Kong Baptist University in 2020. His research interests include General-Purpose GPU Computing and power-efficient computing. He is a recipient of Hong Kong PhD Fellowship.
\end{IEEEbiography}
\vspace{-4.0em}
\begin{IEEEbiography}
	[{\includegraphics[width=1.15in,height=1.35in,clip,keepaspectratio]
		{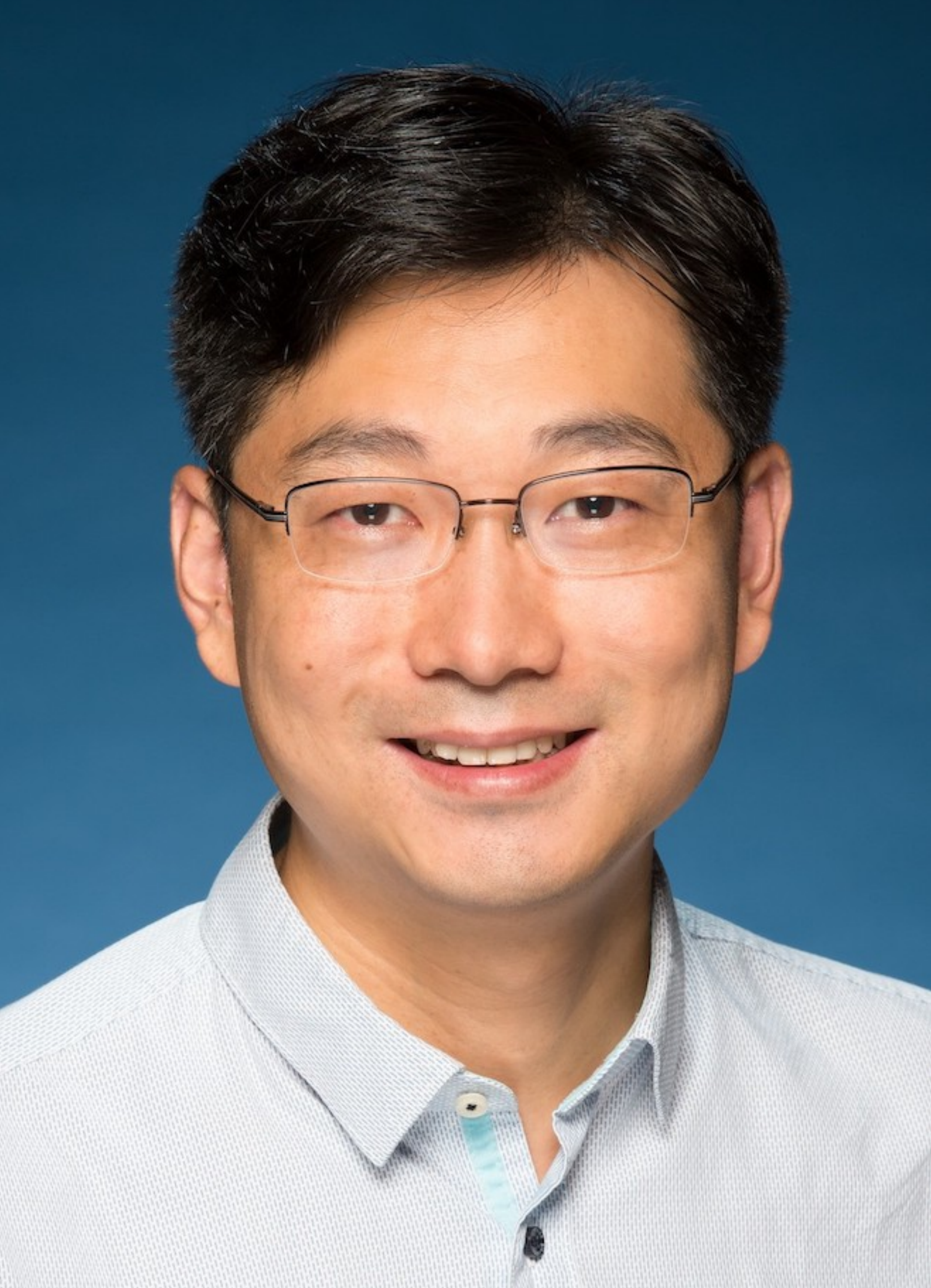}}]{Xiaowen Chu}
	received the B.E. degree in computer science from Tsinghua University, P.R. China, in 1999, and the Ph.D. degree in computer science from The Hong Kong University of Science and Technology in 2003. Currently, he is a full professor in the Department of Computer Science, Hong Kong Baptist University. His research interests include distributed and parallel computing and wireless networks. He is serving as an Associate Editor of IEEE Access and IEEE Internet of Things Journal.
\end{IEEEbiography}
\vspace{-4.0em}
\begin{IEEEbiography}
	[{\includegraphics[width=1in,height=1.25in,clip,keepaspectratio]
		{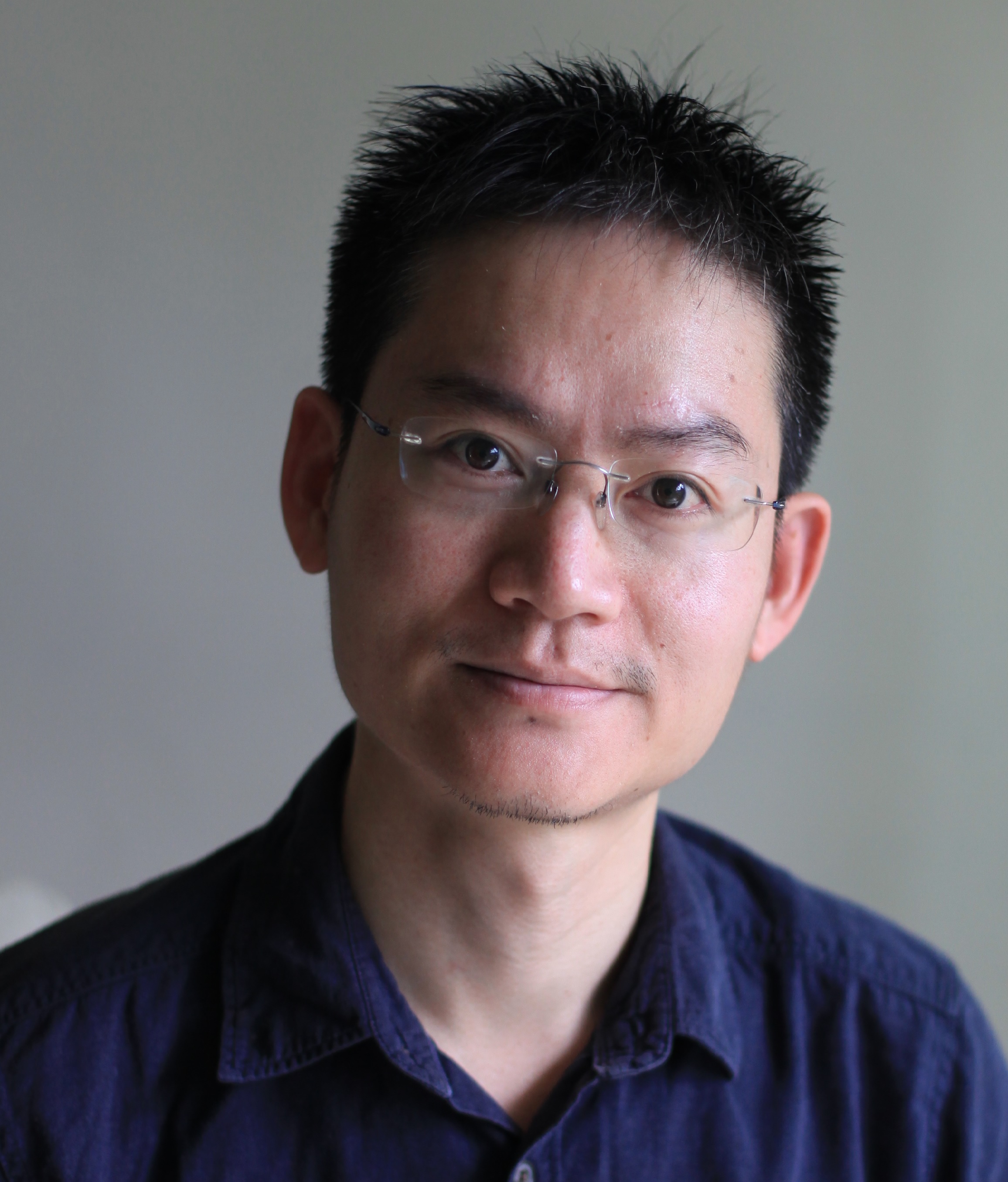}}]{Hai Liu}
	is an Associate Professor with Department of Computing, The Hang Seng University of Hong Kong. Before joining HSUHK, he held several academic posts at University of Ottawa and Hong Kong Baptist University. Dr. Liu received PhD in Computer Science at City University of Hong Kong, and received MSc and BSc in Applied Mathematics at South China University of Technology. His research interest includes wireless networking, cloud computing and algorithm design and analysis. His h-index is 27 according to Google Scholar. He is a member of IEEE.
\end{IEEEbiography}
\vspace{-4.0em}
\begin{IEEEbiography}
	[{\includegraphics[width=1in,height=1.25in,clip,keepaspectratio]
		{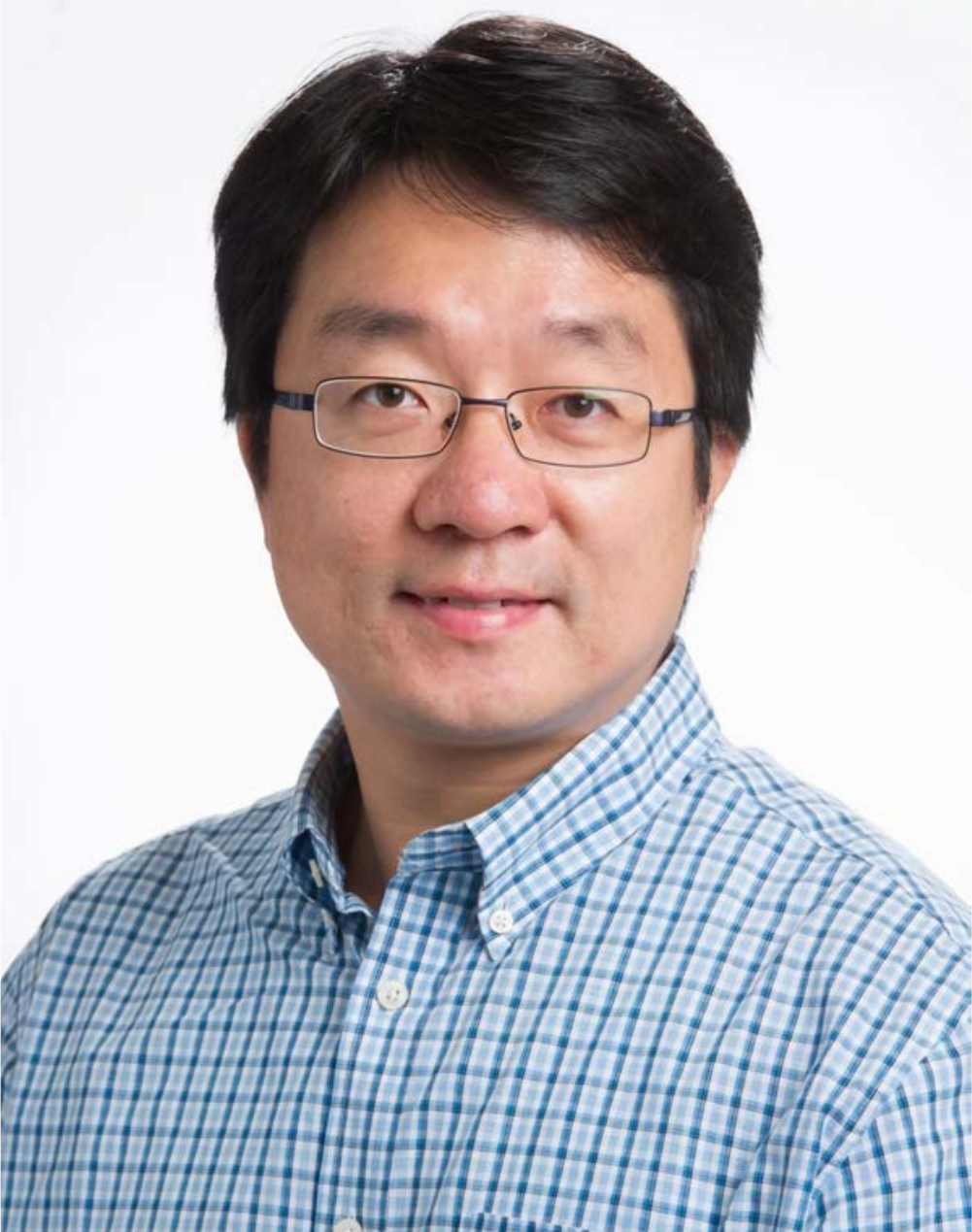}}]{Yiu-Wing Leung}
	received his B.Sc. and Ph.D. degrees from the Chinese University of Hong Kong. He has been working in the Hong Kong Baptist University and now he is Professor of the Computer Science Department and Programme Director of two MSc programmes. His research interests include three major areas: 1) network design, analysis and optimization, 2) Internet and cloud computing, and 3) systems engineering and optimization. He has published more than 50 papers in these areas in various IEEE transactions and journals.
\end{IEEEbiography}
\vspace{-4.0em}
\begin{IEEEbiography}
	[{\includegraphics[width=1in,height=1.25in,clip,keepaspectratio]
		{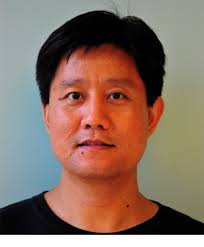}}]{Zongpeng Li}
	received his B.E. degree in Computer Science from Tsinghua University in 1999,
	and his Ph.D. degree from University of Toronto in 2005. He has been with the University of Calgary and then Wuhan University. His research
	interests are in computer networks and cloud
	computing. Zongpeng was named an Edward
	S. Rogers Sr. Scholar in 2004, won the Alberta
	Ingenuity New Faculty Award in 2007, and was
	nominated for the Alfred P. Sloan Research Fellow in 2007. Zongpeng co-authored papers that
	received Best Paper Awards at the following conferences: PAM 2008,
	HotPOST 2012, and ACM e-Energy 2016. Zongpeng received the Department Excellence Award from the Department of Computer Science,
	University of Calgary, the Outstanding Young Computer Science Researcher Prize from the Canadian Association of Computer Science,
	and the Research Excellence Award from the Faculty of Science, University of Calgary.
\end{IEEEbiography}
	
\end{document}